\documentclass[letterpaper,11pt]{article}
\pdfoutput=1 

\usepackage{jheppub}
\usepackage{graphicx}
\usepackage{hyperref}
\usepackage{caption,subcaption} 
\usepackage[section]{placeins}

\newcommand{\comment}[1]{}
\newcommand{\beq}[1]{\begin{equation}\label{#1}}
\newcommand{\eeq}{\end{equation}}
\renewcommand{\O}{\mathcal{O}}
\newcommand{\C}{\mathcal{C}}
\newcommand{\B}{\mathcal{B}}
\newcommand{\del}{\partial}
\newcommand{\Del}{\nabla}
\newcommand{\M}{\mathcal{M}}

\title{Complexity of Scalar Collapse in Anti-de Sitter Spacetime}

\author[a]{Andrew R.~Frey,}
\author[a]{Michael P.~Grehan,}
\author[b,1]{and Manu Srivastava\note{Previous affiliation: 
Department of Physics, Indian Institute of 
Technology Bombay, Mumbai 400076, India} }

\affiliation[a]{Department of Physics and Winnipeg Institute for
Theoretical Physics, University of Winnipeg\\
515 Portage Avenue, Winnipeg, Manitoba R3B 2E9, Canada}

\affiliation[b]{Perimeter Scholars International, Perimeter Institute for
Theoretical Physics,
31 Caroline Street North, Waterloo, Ontario N2L 2Y5, Canada}

\emailAdd{a.frey@uwinnipeg.ca}
\emailAdd{grehan-m@webmail.uwinnipeg.ca}
\emailAdd{msrivastava@perimeterinstitute.ca}

\abstract{We calculate the volume and action forms of holographic 
complexity for the gravitational collapse of scalar field matter in
asymptotically anti-de Sitter spacetime, using numerical methods to
reproduce the geometry responding to the oscillating field over multiple 
crossing times. Like the scalar field pulse, the volume complexity 
oscillates quasiperiodically before horizon formation. It also shows a 
scaling symmetry with the amplitude of the scalar field. The action 
complexity is also quasiperiodic with spikes of increasing amplitude.
}


\begin{document}
\maketitle

\section{Introduction}\label{s:intro}

Recent years have seen a rapid growth of interest in (circuit) complexity
as an information theoretic quantity of relevance to fundamental physics,
following the foundational work of 
\cite{arXiv:1402.5674,arXiv:1406.2678,arXiv:1509.07876,arXiv:1512.04993}.
In quantum mechanics and quantum field theory, complexity is a measure of 
distance of the state in question from a reference state 
\cite{quant-ph/0502070,quant-ph/0603161,quant-ph/0701004,arXiv:1707.08570,arXiv:1801.07620,arXiv:1803.10638}
representing the minimal amount of computational processing needed to 
generate that state (i.e., the size of the optimal circuit required in a quantum
computer using a specified set of basis gates). 
Therefore, like entropy represents the information content of 
a system, the time derivative of complexity represents processing speed.

Also like entropy, the complexity of states in field theories with 
gravity duals should have a holographic interpretation
in terms of geometry. In the AdS/CFT correspondence,
there are in fact two proposals for a holographic
definition of complexity for the state on a spacelike slice $\Sigma$ of
the conformal boundary of anti-de Sitter spacetime (AdS). The 
first is ``$\C=V$'' complexity
\cite{arXiv:1402.5674,arXiv:1406.2678}, which 
assigns the complexity as $\C_V = \max [V(\B)/G\ell]$, where the maximum is 
over spacelike bulk surfaces $\B$ with boundary on $\Sigma$, 
$G$ is the Newton constant of the AdS spacetime, and $\ell$ is some length 
scale depending on the asymptotically AdS geometry (typically either the
AdS curvature scale or the Schwarzschild radius of a small black hole).
The other proposal, ``$\C=A$,'' states that the complexity of the state on
$\Sigma$ is $\C_A=S_{WDW}/\pi$ (in units with $\hbar=1$), where $S_{WDW}$
is the action of the Wheeler--DeWitt (WDW) patch of $\Sigma$ in the bulk
\cite{arXiv:1509.07876,arXiv:1512.04993}.
The WDW patch is the union of spacelike bulk surfaces with boundary on
$\Sigma$, so the WDW patch is bounded by past- and future-directed 
lightlike surfaces
emitted from $\Sigma$ into the bulk. $S_{WDW}$ includes boundary terms as well
as joint terms on $\Sigma$.

In both proposals, the complexity of eternal AdS black holes exhibits
linear growth in time, as expected for a thermal equilibrium state
(see \cite{arXiv:1406.2678,arXiv:1509.07876} for the initial calculations).
Of course, the contribution of time evolution of the spacetime background
to time dependence of complexity is an important question; 
\cite{arXiv:1408.2823,arXiv:1711.02668,arXiv:1802.06740,arXiv:1803.11162,arXiv:1804.07410,Lezgi:2021qog} 
have studied 
complexity in black hole formation due to the collapse of a thin shell of
null matter (i.e., the AdS-Vaidya geometry) as a probe both of  
complexity during thermalization of the dual CFT and of both proposals for
holographic complexity. As summarized in \cite{arXiv:1804.07410}, 
the complexity of the collapsing geometry approaches the linear growth
rate of eternal black holes at late times following transient behavior
in the early stages of the shell's collapse.

Our purpose is to examine the transient early time behavior of complexity 
for the collapse of a smooth distribution of matter rather than a thin shell.
There is a critical difference between these two cases in asymptotically
AdS spacetime with the conformal boundary $\mathbb{R}\times S^{d-1}$ of AdS in 
global coordinates (henceforth, ``asymptotically global AdS''): while
thin shells always form an apparent horizon promptly, a smooth distribution 
of matter may collapse, disperse, reflect from the conformal boundary,
and collapse again many times before finally forming a horizon, as argued
initially for massless scalars with spherical symmetry in 
\cite{arXiv:1104.3702,arXiv:1106.2339,arXiv:1108.4539,arXiv:1110.5823}.
For a fixed profile of initial data for the scalar field, lower initial 
amplitudes take a longer time to form a black hole, but instability to
eventual horizon
formation seems to be generic; however, there is a so-called ``island
of stability'' for initial profiles with characteristic width near the AdS
scale with quasiperiodic scalar and gravitational evolution 
\cite{arXiv:1304.4166,arXiv:1307.2875,arXiv:1308.1235,Evnin:2021buq}.
Because unstable initial data also evolves quasiperiodically prior to 
horizon formation, determining whether a given set of initial data is 
stable or unstable at low amplitudes is a topic of interest in the literature.
While massless scalars are better studied,
massive scalars exhibit similar behavior
\cite{arXiv:1504.05203,arXiv:1508.02709,arXiv:1711.00454}.

As a step toward understanding complexity in time-dependent asymptotically
global AdS spacetimes with smooth matter distributions,
we calculate volume and action complexity at early times in
the gravitational collapse of spherically symmetric scalar fields. 
One question of interest is whether complexity evolves quasiperiodically 
or if it demonstrates a trend over oscillations of the scalar field packet,
which we will refer to as ``ratcheting'' behavior. We might expect
ratcheting as an indication that the scalar field undergoes gravitational
focusing throughout its oscillation, which eventually leads to horizon
formation.

We begin with a review of scalar field collapse in AdS, including 
our numerical techniques, in section \ref{s:collapse}. We then present
the volume complexity for a variety of initial data in section \ref{s:volume}
and action complexity in \ref{s:action}. We conclude with a discussion 
of our results and their connection to other studies of holographic 
complexity in section \ref{s:discussion}.

\section{Spherically Symmetric Scalar Field Collapse in AdS}\label{s:collapse}

\subsection{Review}\label{s:review}
We wish to describe the complexity of gravitational collapse in 
asymptotically global AdS spacetime driven by scalar field matter. 
For convenience in solving the coupled Einstein and Klein-Gordon equations
\cite{arXiv:1608.05402}, 
we normalize the scalar so the bulk action takes the form
\beq{action-bulk}
S = \frac{1}{16\pi G} \int d^{d+1}x \sqrt{-g} \left[ \vphantom{\frac 12}
R + d(d-1) - (d-1) \partial_\mu \phi \partial^\mu \phi 
-(d-1)\mu^2\phi^2\right]
\eeq 
in AdS units. For simplicity, we assume spherical symmetry in the collapse
process and write the backreacted geometry in Schwarzschild-like coordinates 
\beq{metric}
ds^2  =\sec^2x\left[ -A(t,x)e^{-2\delta(t,x)} dt^2 + A(t,x)^{-1} dx^2 + 
(\sin^2x)d\Omega^2_{d-1} \right] .
\eeq
The metric functions satisfy boundary conditions $A(t,0)=A(t,\pi/2)=1$
and $\delta(t,0)=0$; with this choice, the coordinate $t$ represents proper
time at the origin, while the proper time $\tau$ of the conformal boundary
satisfies $d\tau/dt =\exp[-\delta(t,\pi/2)]$. 

In terms of the scalar canonical momentum $\Pi$ and an auxiliary field
$\Phi =\del_x\phi$, the Klein-Gordon equation in first order form is
\beq{kg}
\del_t\phi=Ae^{-\delta}\Pi,\quad\del_t\Phi=\del_x\left(Ae^{-\delta}\Pi\right),\quad
\del_t\Pi=\frac{\del_x(Ae^{-\delta}\tan^{d-1}(x)\Phi)}{\tan^{d-1}(x)}-
\frac{e^{-\delta}\mu^2\phi}{\cos^2(x)}.\eeq
Due to spherical symmetry,
the Einstein equations are constraints that determine $A$ and $\delta$
in terms of the scalar field and its derivatives. These are
\beq{constraints}
\del_x\delta=-\sin(x)\cos(x)(\Pi^2+\Phi^2),\quad
A=1-2\frac{\sin^2(x)}{(d-1)}\frac{M}{\tan^{d}(x)},\eeq
where
\beq{mass}
\del_x M = \left(\tan(x)\right)^{d-1}\left[A\frac{\left(\Pi^2+\Phi^2\right)}{2}+
\frac{\mu^2\phi^2}{2\cos^2(x)}\right]\eeq
gives the mass function $M$. When we allow only normalizable modes of the
scalar field, $\M=M(t,\pi/2)$ is the conserved mass of the 
spacetime (above the empty AdS value); 
for initial conditions that form a black hole, $\M$ is the
eventual mass of the black hole.

Because $\M$ is conserved, the energy in the scalar field cannot dissipate,
even if it does not form a horizon promptly. After failing to form a horizon,
a scalar wave can travel to the conformal boundary, reflect, and recollapse.
Therefore, well-defined pulses may form a horizon any time the pulse 
approaches the origin, which occurs at time intervals of 
$\pi$.\footnote{slightly modified by time dilation effects} For massive
scalar fields, the pulses do not reach the conformal boundary but do oscillate
around the origin and can form a black hole at a late time. For many 
initial field profiles, any amplitude $a$ for the initial data 
eventually leads to horizon formation. At small amplitude, a perturbative
analysis guarantees that the horizon forms at times $\sim\O(a^{-2})$ (for any
approximate definition of horizon formation). However, when the typical length
scale of the initial field profile is similar to the AdS scale, scalar
fields below a critical initial amplitude appear to oscillate quasiperiodically
without forming a black hole. The location of this ``island of stability'' 
and the physics underlying the difference between unstable (horizon forming)
and stable (quasiperiodic) solutions have been the subject of much study 
(see \cite{arXiv:1403.6471,arXiv:1410.2631,arXiv:1506.07907,arXiv:1412.3249,arXiv:1412.4761,arXiv:1407.6273,arXiv:1304.4166,arXiv:1307.2875,arXiv:1504.05203,arXiv:1508.02709,arXiv:1711.00454}
for example).

It is also possible to consider the injection of a scalar field pulse at the
conformal boundary, similar to the Vaidya solution, which corresponds to
allowing non-normalizable modes of the scalar field.\footnote{In this
case, $\M$ is not conserved as long as $\Pi(t,\pi/2)\neq 0$ (during the
injection of the pulse), but it is afterwards.} This scenario is less 
well-understood than the case of an initial scalar profile, but see
\cite{arXiv:1612.07701,arXiv:1712.07637,arXiv:1912.07143}. 
Due to computational requirements discussed below, we will consider the
complexity in black hole formation with nontrivial initial conditions
rather than nontrivial boundary conditions for the scalar field.

\subsection{Numerical Solutions}\label{s:numerics}

We use a 4th-order Runge-Kutta time stepper and spatial integrator to solve 
the Klein-Gordon and Einstein equations over a grid of $2^n+1$ spatial points
(see \cite{arXiv:1410.1869,arXiv:1508.02709} for a more detailed description 
of the code, which is similar to that of \cite{arXiv:1308.1235}). 
While studies of the stability of AdS vs horizon formation require long-time
numerical stability, necessitating higher spatial resolution, we currently
want to reproduce the spacetime in sufficient detail to calculate maximal
surface volumes and WDW patch actions. We therefore need to save both 
functions $A(t,x),\delta(t,x)$ to disk frequently; for storage reasons, we
will study numerical evolutions that are numerically convergent for a 
low value $n=10$. We have carried out convergence tests as described in
\cite{arXiv:1508.02709}
to verify that all the evolutions studied are in fact convergent.

Strictly speaking, a horizon forms only at infinite time in the coordinates
we have chosen (this corresponds to the exponential approach of the dual
field theory to equilibrium). However, we can denote approximate horizon
formation at position $x_H$ and time $t_H$ by the first point such that 
$A(x_H,t_H)\leq 2^{7-n}$. Our solver then terminates at time $t_H$, so we can
only study the spacetime for $0\leq t\leq t_H$.

\begin{figure}[!t]
\centering
\begin{subfigure}[t]{0.47\textwidth}
\includegraphics[width=\textwidth]{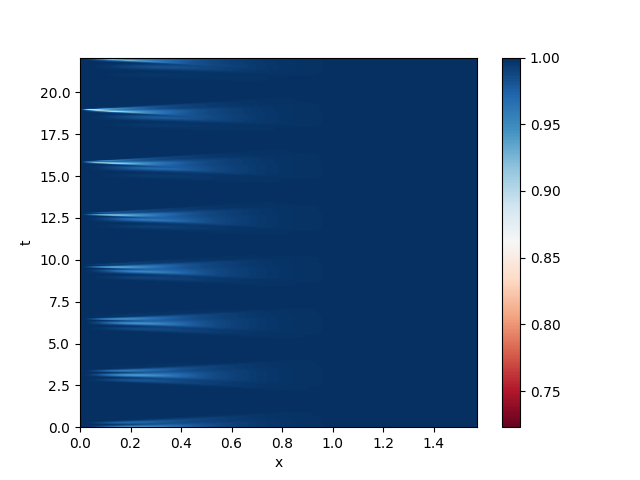}
\caption{$A(t,x)$}\label{f:Aspacetime}
\end{subfigure}
\begin{subfigure}[t]{0.47\textwidth}
\includegraphics[width=\textwidth]{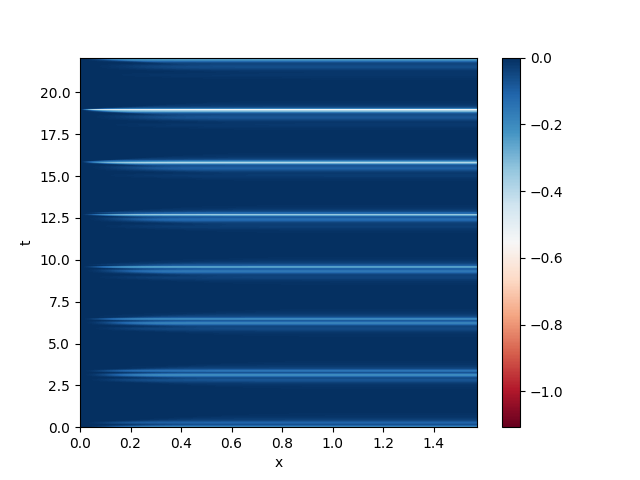}
\caption{$\delta(t,x)$}\label{f:deltaspacetime}
\end{subfigure}
\caption{Metric functions for collapse of a massless scalar in 
AdS$_5$ with initial width $\sigma=0.25$ and amplitude $a=3.87$}
\label{f:spacetime}
\end{figure}

The closest equivalent to the thin shell Vaidya spacetime is to inject
a scalar field pulse at the conformal boundary via time-dependent
boundary conditions for $\phi$. However, based on the experience of
\cite{deppefreyhoult}, numerical convergence fails even at relatively 
short times for
$n<15$. Therefore, we instead require that $\phi(t,\pi/2)\to 0$ (with the
appropriate fall off for normalizable modes) and take initial data such
that the canonical momentum is Gaussian in the radius,
\beq{PiGaussianID}
\Pi(t=0,x)=a\exp\left(-\frac{\tan^2(x)}{\sigma^2}\right),
\quad\phi(t=0,x)=0.
\eeq

We use linear interpolation
to find values of $A(t,x),\delta(t,x)$ (as well as scalar field quantities)
between spatial grid points and time steps for use in calculating the
complexities as described below. Figure \ref{f:spacetime} shows the metric
functions $A(t,x),\delta(t,x)$ for an example set of the initial data that
forms a horizon after the scalar pulse reflects from the conformal boundary
seven times. Each time the pulse approaches the origin (when $t$ is a multiple
of $\pi$), the minimum value of $A$ gets smaller and $\delta$ becomes more
negative (for all $x$); this effect intensifies each time. From this point
of view, the pulse focuses as time goes on, which suggests that the
complexity may ratchet to progressively larger (or even smaller) values
during the pre-collapse phase of scalar field evolution.

\section{Complexity as Volume}\label{s:volume}

\subsection{Set-Up}\label{s:CVsetup}
We recall that the volume complexity of the boundary CFT state on a spacelike
slice $\Sigma$ is given by $\C_V = \max [V(\B)/G\ell]$, where the maximum is 
taken over surfaces $\B$ such that $\del\B=\Sigma$ and $\ell$ is a specified
length scale. In most cases, $\ell$ is taken to be the AdS curvature radius
($=1$ in our units), but it is $\ell=r_+$ the Schwarzschild radius for
small black holes in AdS. In this way, $\ell$ is the maximal time to fall
from the horizon to the final cylinder in either a large or small black hole
\cite{arXiv:1807.02186}. While it is unclear how to generalize the relationship 
described by this prescription to a dynamical setting, we consider initial
conditions that eventually lead to small black holes and therefore 
(for most of our discussion) take
$\ell=r_{+,f}$, the Schwarzschild radius of the final equilibrium black hole
after all the initial energy falls behind the horizon. However, we discuss
the alternate prescription taking $\ell=1$ below as well.

Because the asymptotically AdS$_{d+1}$ metric (\ref{metric}) is spherically
symmetric, the maximal volume surface is a surface of revolution obtained by
rotating the curve $(t(\lambda),x(\lambda))$ around the $S^{d-1}$ with 
the boundary condition $x(\lambda\to\infty)\to\pi/2$. The value
of $t(\lambda\to\infty)$ is the time at which we measure the complexity.
The volume of any such surface is 
\beq{volume1}
V =  \frac{2\pi^{(d-1)/2}}{\Gamma((d-1)/2))} \int d\lambda\, 
\tan^{(d-1)} (x) \sec(x) 
\sqrt{ (x^\prime)^2 A^{-1} - (t^\prime)^2 A e^{-2\delta}  }  ,
\eeq
where a prime indicates the derivative with respect to $\lambda$. 
We can choose the parameter $\lambda$ to set the integrand of (\ref{volume1})
equal to unity, in which case the Euler-Lagrange equations have a 
first-order form
\begin{align}
(P^{x})^\prime & = -(P^x)^2 \left( \frac{A_x}{2\Omega} \right) 
- (P^t)^2 \left( \frac{e^{2\delta} }{ 2A^2 \Omega } \right) (A_x - 2A\delta_x ) 
+ (d-1)\cot(x) + d\tan(x) ,\label{Pxprime}\\
(P^t)^\prime & = -(P^x)^2 \left( \frac{A_t}{2\Omega} \right) 
- (P^t)^2 \left( \frac{e^{2\delta} }{2A^2 \Omega } \right) (A_t - 2A\delta_t ) ,
\label{Ptprime}\\
x^\prime & = \left( \frac{A}{\Omega} \right) P^x ,\quad\textnormal{and}\quad
t^\prime = - \left( \frac{e^{2\delta}}{A\Omega} \right) P^t ,\label{xtprime}
\end{align}
where subscript $x,t$ denote partial derivatives and 
$\Omega\equiv \tan^{2(d-1)}x\sec^2 x$. With the given choice of parameter,
\beq{parameterization}
\sqrt{\frac{A(P^x)^2 -A^{-1}e^{2\delta} (P^t)^2}{\Omega}} = 1 .
\eeq

For the surface to be smooth at the origin, $dt/dx=0$, which implies 
$e^{2\delta}P^t/A^2P^x\to 0$ as $x\to 0$. Then our choice of parameter implies
$P^x\approx \sqrt{\Omega/A}$ and $x'\approx\sqrt{A/\Omega}$, 
while the Einstein equation constraints imply that
$A\approx 1+a(t)x^2$ near the origin. Therefore, to the first subleading order,
\beq{lambdainit}
\lambda = \frac{1}{d}x^d + 3 \left( \frac{d-1/2}{d+2} \right) x^{d+2} + 
\frac{1}{2(d+2)}x^d\left( A(t,x) - 1\right) .
\eeq
Meanwhile, equation (\ref{Ptprime}) yields $dP^t/dx\approx -A_t/2$, so 
\beq{Ptinit}
P^t = -\frac{1}{2(d+2)} A_t (x, t) x^d
\eeq
to order $x^{d+2}$. To this order, $t$ is constant.
We can therefore solve 
(\ref{Pxprime},\ref{Ptprime},\ref{xtprime}) starting from a given $t_0$ and
$x_0\ll 1$ using (\ref{lambdainit},\ref{Ptinit}) as initial values at $x_0$.

Near the conformal boundary, $A_x,A_t\to 0$, while $\delta\to\delta(t)$.
As a result, we have a self-consistent solution with $(P^x)'\approx d\tan(x)$, 
or $P^x\approx \sqrt{\Omega}$ to leading order. In particular, this 
means that $x'\approx 1/\sqrt{\Omega}$, so the surface requires an infinite
parameter $\lambda$ to reach the boundary. We therefore cut off the surface
at $x=\pi/2-\epsilon$. The value of $t$ at this point is not quite the
value at the boundary; however, the difference is $\O(\epsilon^{d+1})$, so we
neglect it.

\subsection{Methods and Results}\label{s:CVresults}

\begin{figure}[t]
\centering
\begin{subfigure}[t]{0.47\textwidth}
\includegraphics[width=\textwidth]{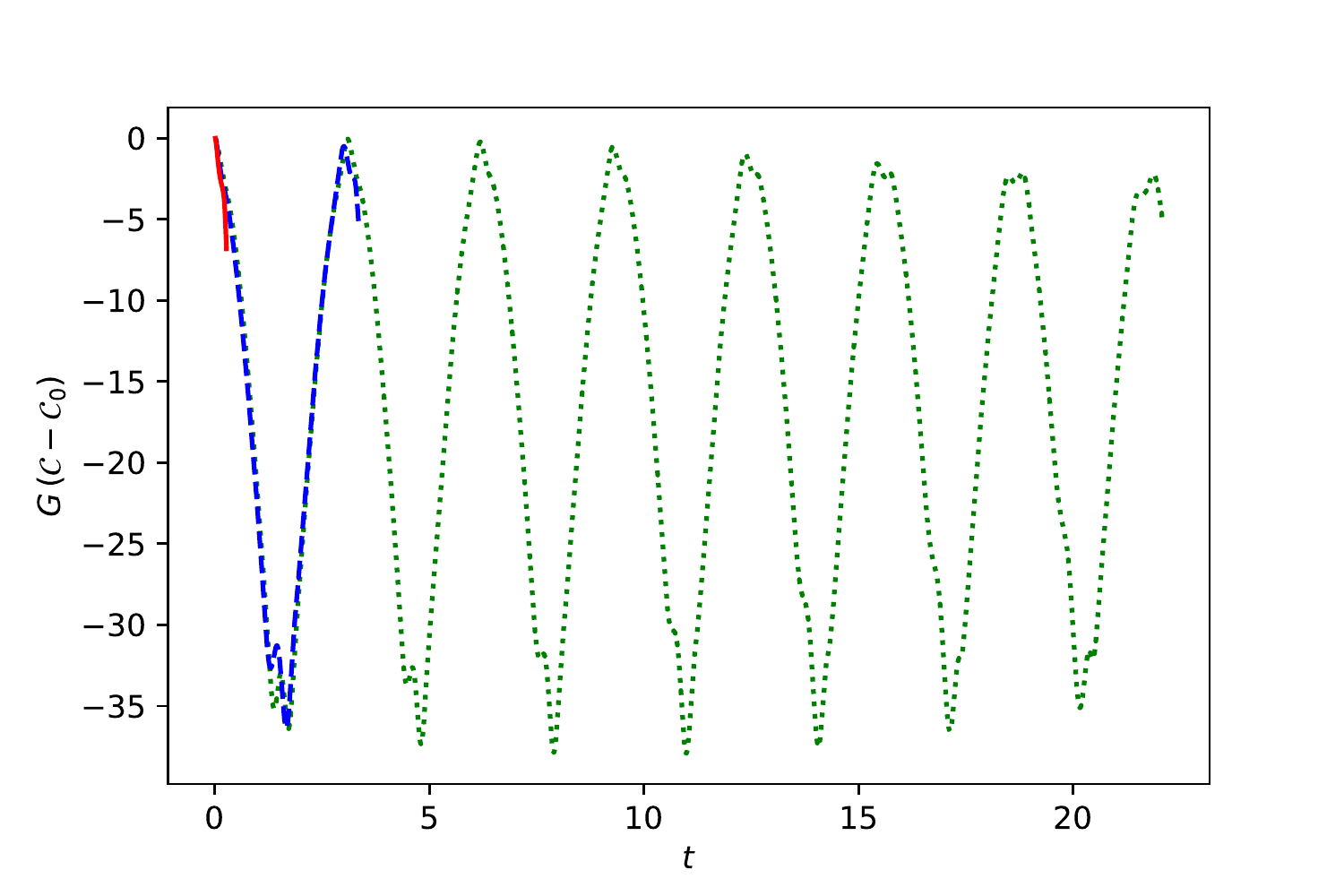}
\caption{AdS$_5$, $\mu=0$, $\sigma=0.25$, $a=8.28,6.39,3.87$}
\label{f:CVt-m0w025d5}\end{subfigure}
\begin{subfigure}[t]{0.47\textwidth}
\includegraphics[width=\textwidth]{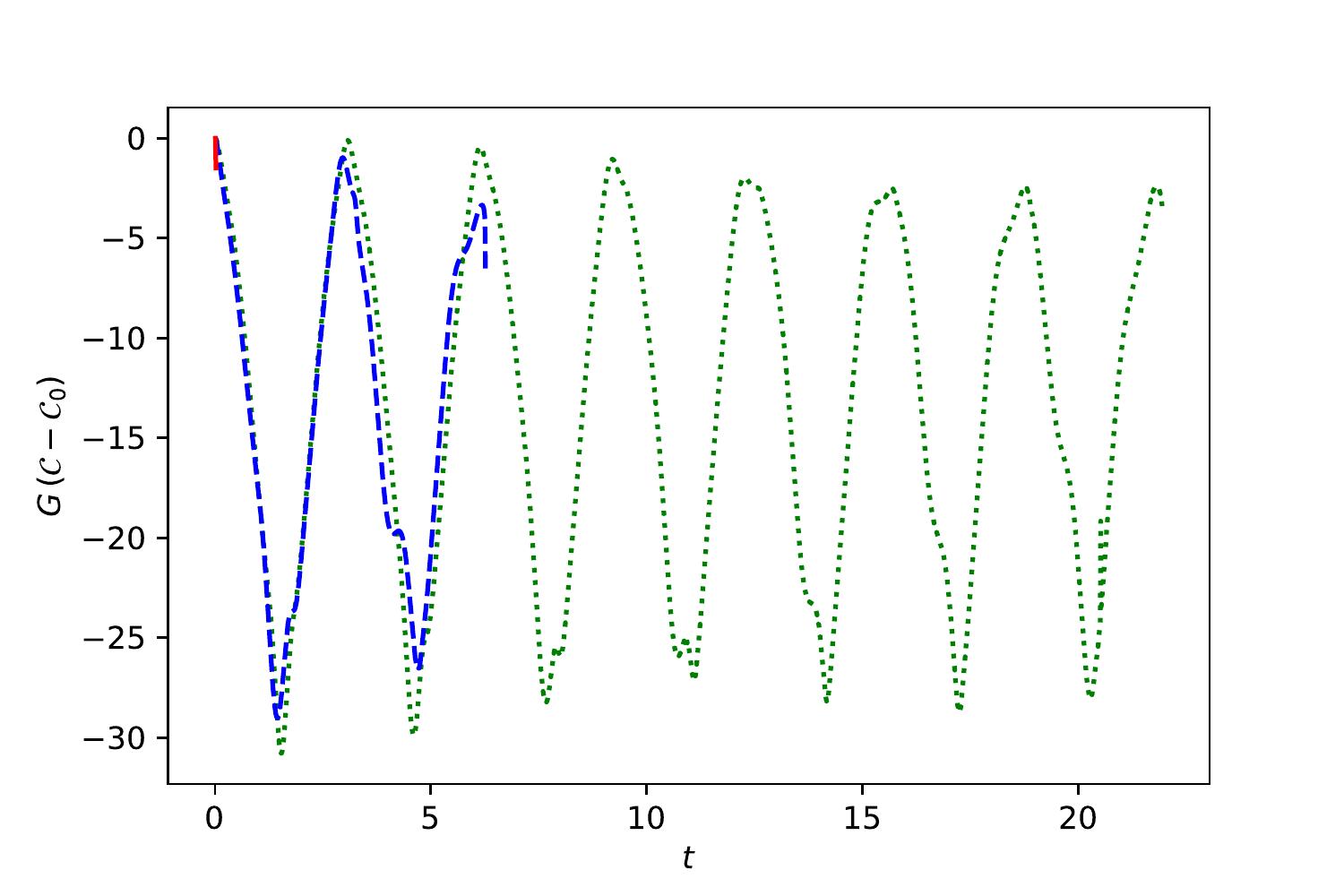}
\caption{AdS$_4$, $\mu=0$, $\sigma=0.25$, $a=15,8.0,5.0$}
\label{f:CVt-m0w025d4}\end{subfigure}
\begin{subfigure}[t]{0.47\textwidth}
\includegraphics[width=\textwidth]{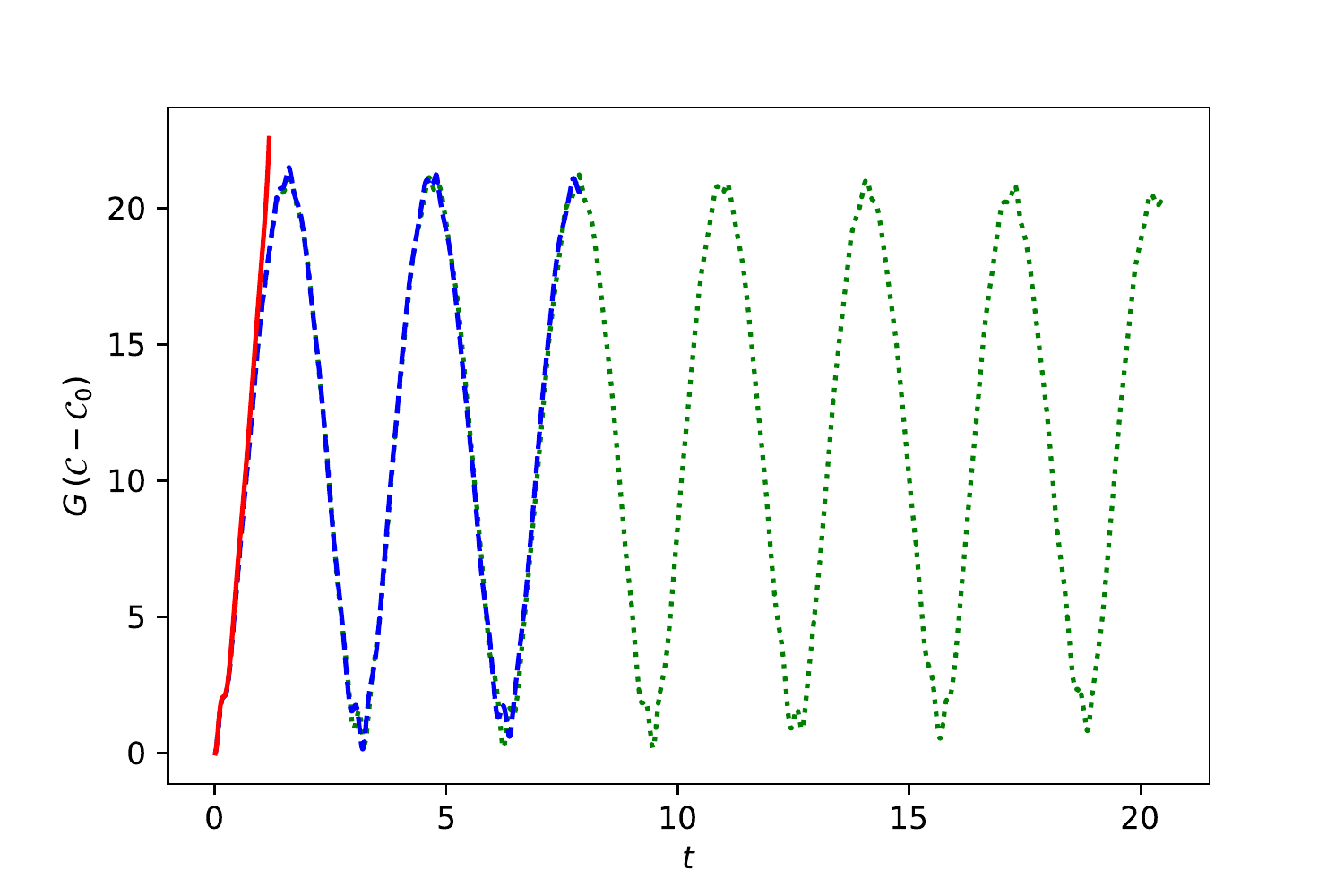}
\caption{AdS$_5$, $\mu=5$, $\sigma=2$, $a=0.38,0.12,0.10$}
\label{f:CVt-m5w2d5}\end{subfigure}
\begin{subfigure}[t]{0.47\textwidth}
\includegraphics[width=\textwidth]{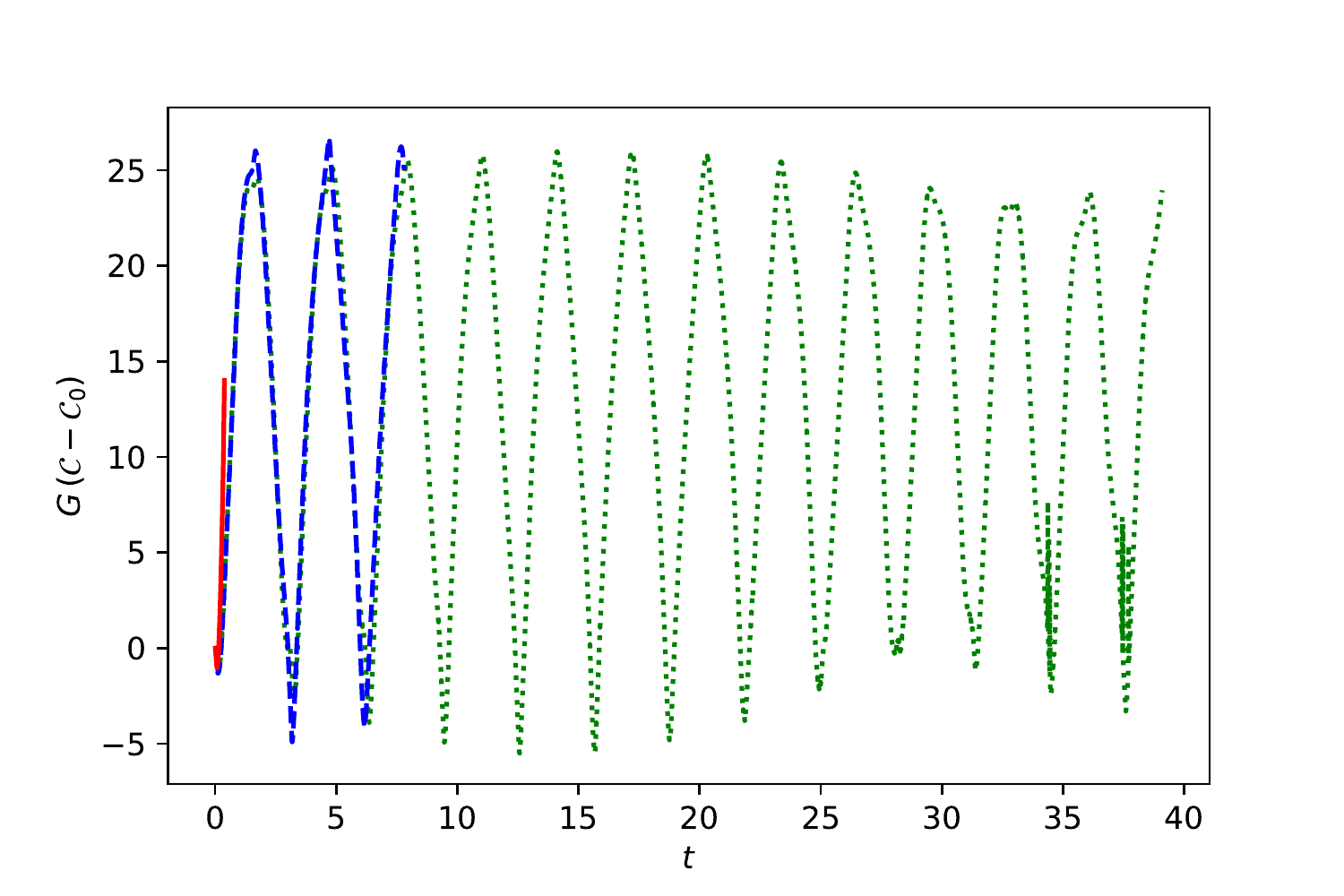}
\caption{AdS$_4$, $\mu=0$, $\sigma=8$, $a=1.00,0.14,0.10$}
\label{f:CVt-m0w8d4}\end{subfigure}
\caption{Volume complexity for listed initial data in AdS$_4$ and AdS$_5$.
Curve for the largest amplitude in each figure is solid red, middle is 
dashed blue, smallest is dotted green.}
\label{f:CVexamples}
\end{figure}

\begin{figure}[t]
\centering
\begin{subfigure}[t]{0.47\textwidth}
\includegraphics[width=\textwidth]{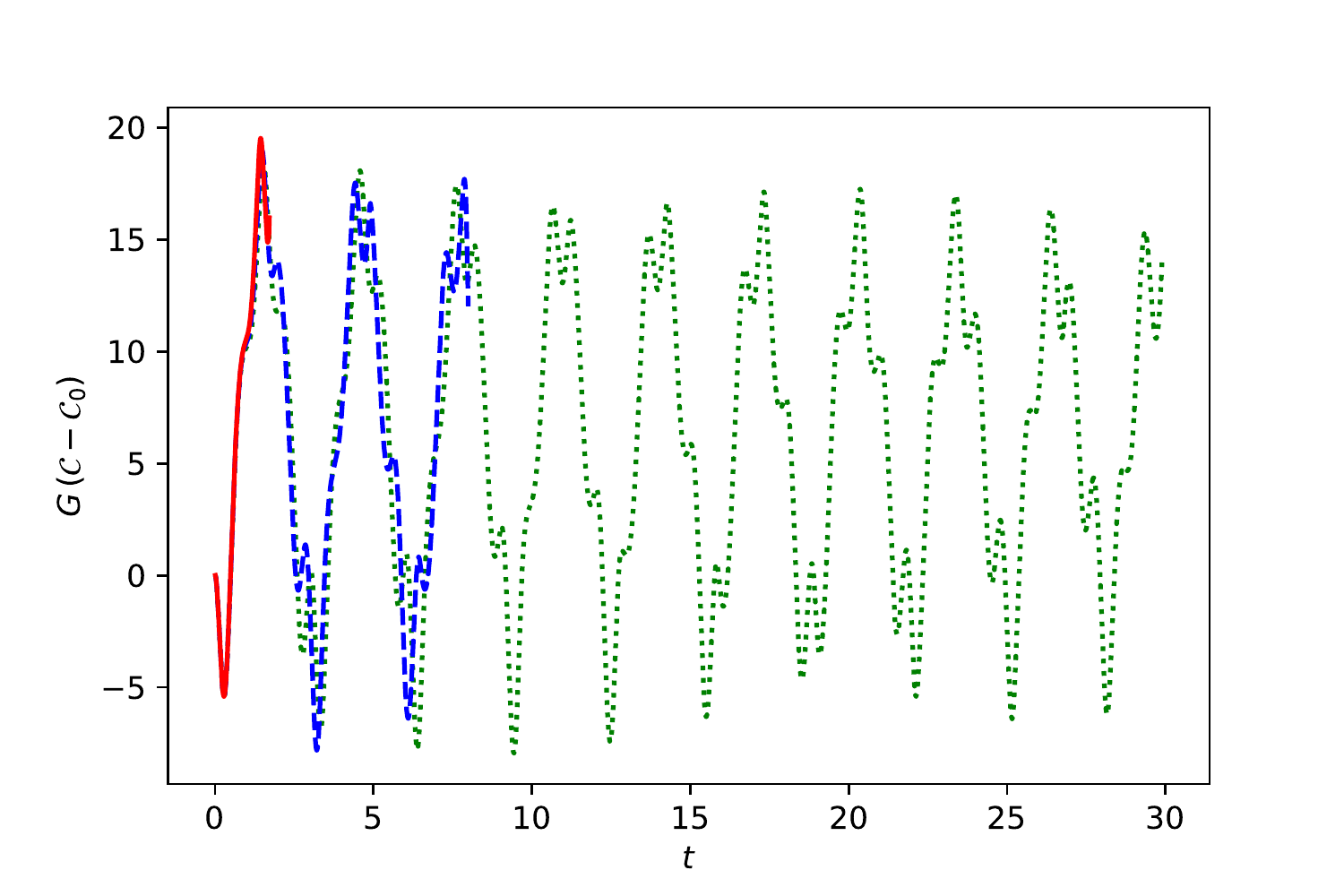}
\caption{$\C$ vs $t$: $\sigma=2.1$, $a=0.28,0.24,0.18$}
\label{f:CVt-m0w21d5}\end{subfigure}
\begin{subfigure}[t]{0.47\textwidth}
\includegraphics[width=\textwidth]{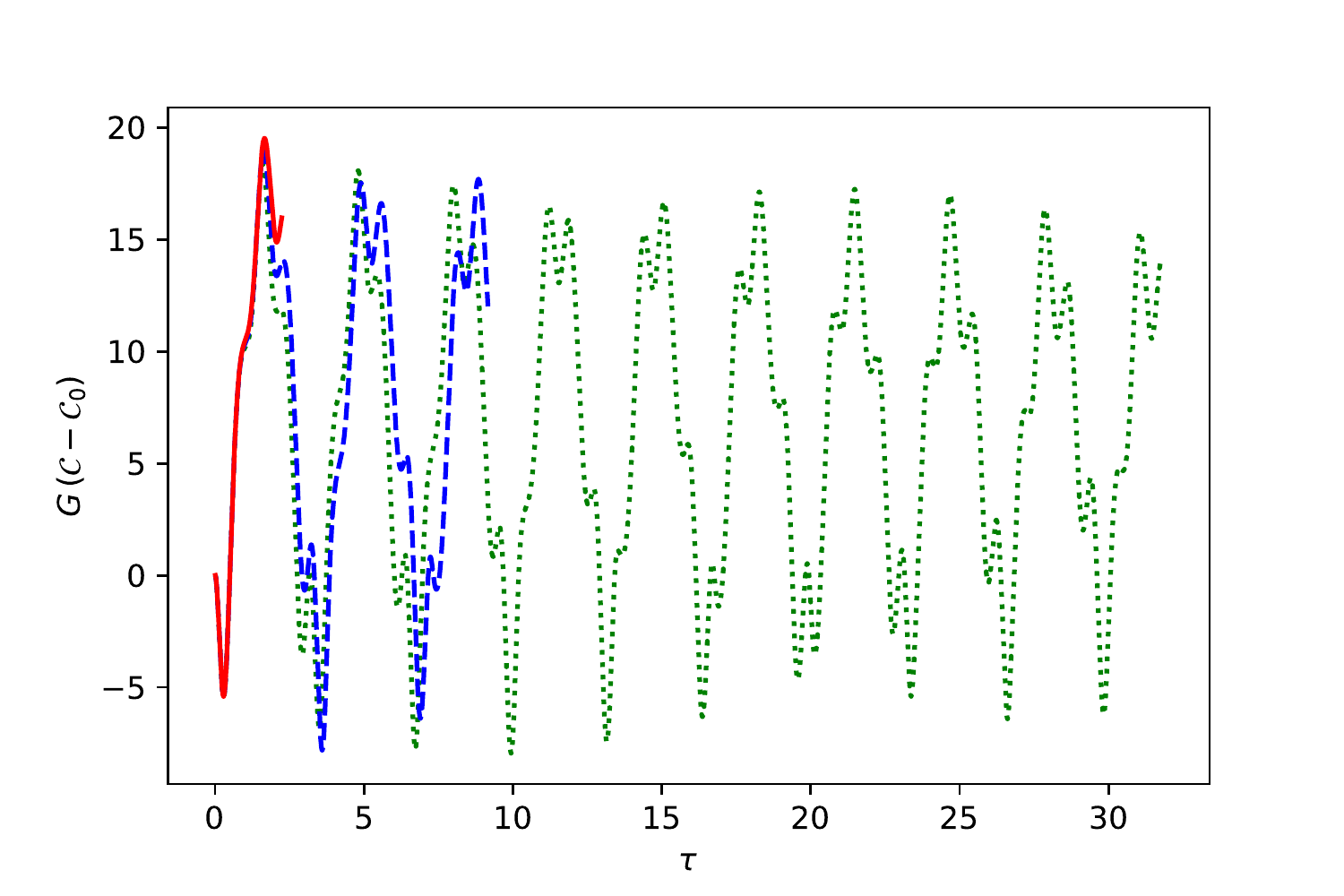}
\caption{$\C$ vs $\tau$: $\sigma=2.1$, $a=0.28,0.24,0.18$}
\label{f:CVtau-m0w21d5}\end{subfigure}
\begin{subfigure}[t]{0.47\textwidth}
\includegraphics[width=\textwidth]{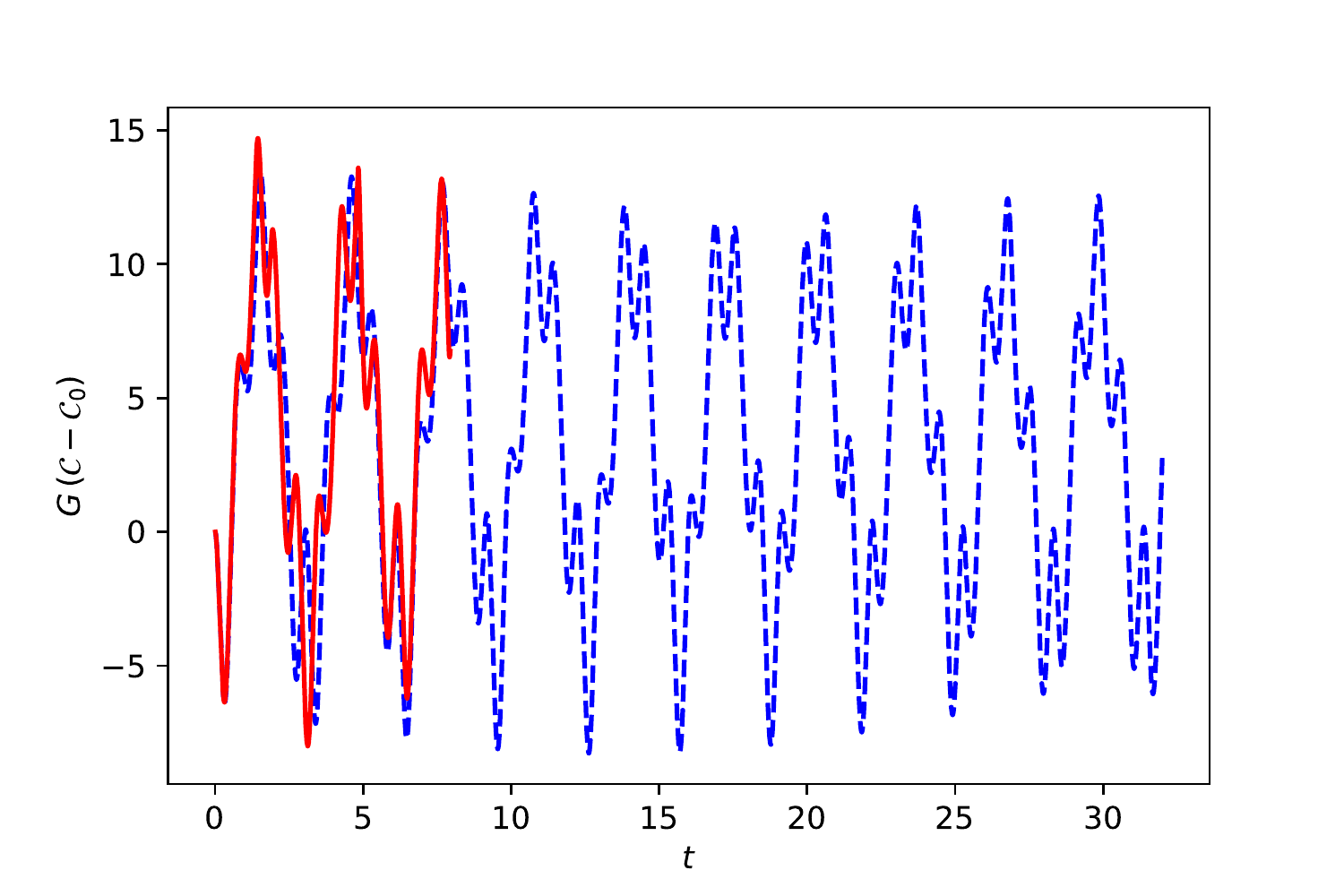}
\caption{$\C$ vs $t$: $\sigma=1.7$, $a=0.38,0.22$}
\label{f:CVt-m0w17d5}\end{subfigure}
\begin{subfigure}[t]{0.47\textwidth}
\includegraphics[width=\textwidth]{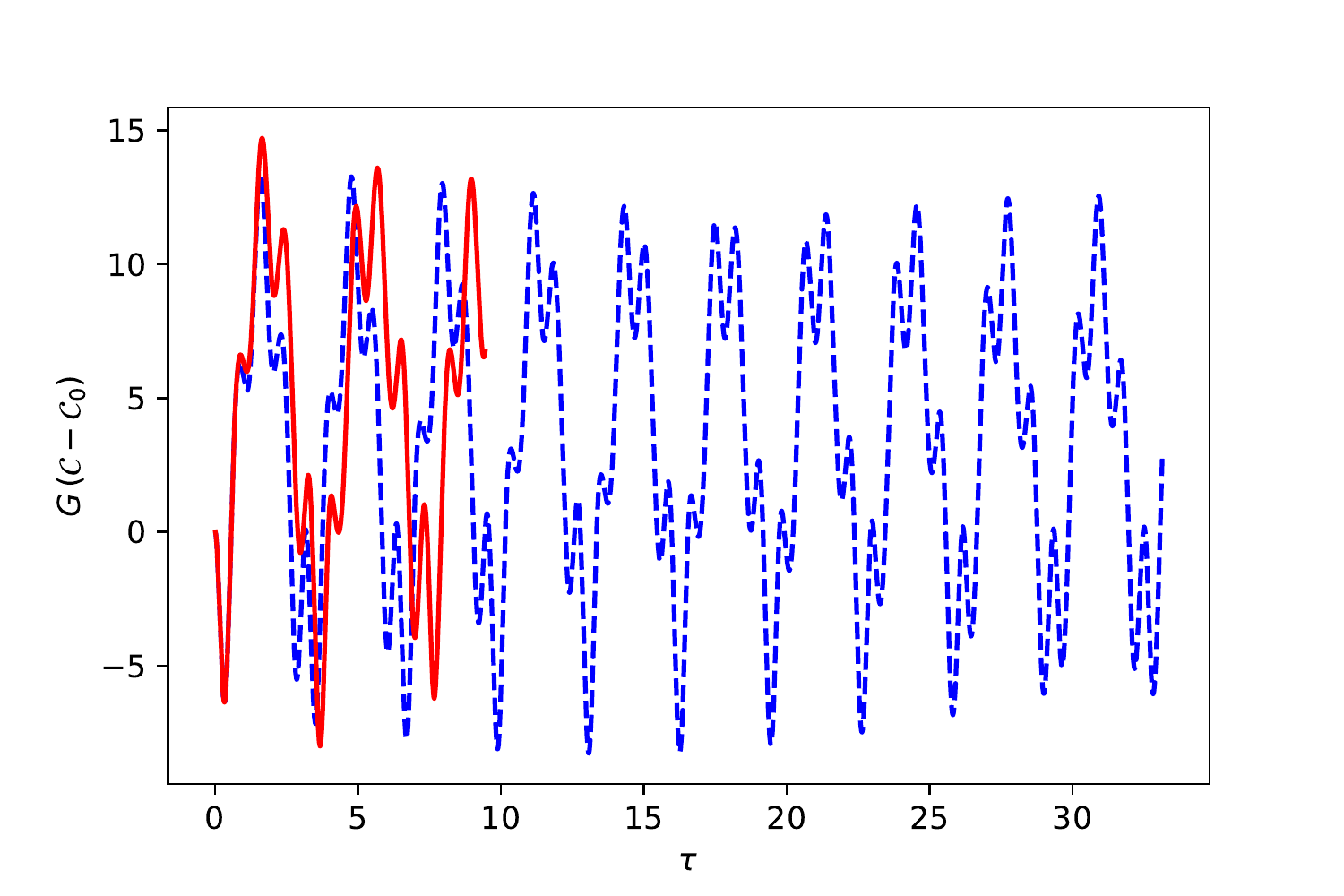}
\caption{$\C$ vs $\tau$: $\sigma=1.7$, $a=0.38,0.22$}
\label{f:CVtau-m0w17d5}\end{subfigure}
\caption{Volume complexity vs $t$ and $\tau$ for two sets of initial data.
Both are massless scalars in AdS$_5$. Curve for the largest amplitude in each 
figure is solid red, next is dashed blue, and next is dotted green.}
\label{f:CVvstandtau}
\end{figure}

To find the maximal volume, we integrate the Euler-Lagrange equations 
using the odeint function of the Python scipy library, reading in the 
outputs of the scalar field collapse simulations. We interpolate 
those functions as described in our discussion of figure \ref{f:spacetime}
at the end of section \ref{s:numerics};
we find partial derivatives of $A,\delta$ through differencing. 

We determine the volume by finding the value of $\lambda$ where $x=1.55$
at a time given by the value of $t$ on the surface at $x=1.55$, which can
alternately be given in terms of the conformal boundary time $\tau$ at that
same point. The complexity at this time is 
\begin{align}
G\C = \left( \frac{2\pi^{(d-1)/2}}{\Gamma(\frac{d-1}{2})}  \right)
\left( \frac{4\pi}{(d-1)\M}\right)\lambda(x=1.55) = 
\left( \frac{4\pi}{(d-1)\M}\right) V,
\end{align}
where the mass gives the final equilibrium black hole Schwarzschild radius
with our normalization for the scalar field. To isolate the time dependence
of the complexity, we subtract $\C_0$, the value at $t=\tau=0$. That is, while 
we regulated the divergent volume by cutting it off 
at radial coordinate $x=1.55$, we display the time dependence by 
renormalizing away the initial value of the volume, which is just the
empty AdS value plus an $\O(a^2)$ correction. We also note that the factor of
$\M$ in the denominator appears because we take $\ell=r_+$ in the definition
of complexity. Since the conserved mass $\M$ is proportional to $a^2$ at
small amplitudes (due to backreaction, the relationship is not so clean at
larger amplitudes), converting to the prescription where
$\ell$ is the AdS radius roughly scales $\C$ by $a^{2}$.

Figure \ref{f:CVexamples} shows examples of volume complexity curves as a 
function of origin time $t$ for a variety of initial data as listed in the 
captions. Each curve terminates when the evolution satisfies our approximate
criterion for horizon formation. 
There are several features in common for all the complexity curves,
including for a massive scalar field.
First, the complexity appears quasiperiodic without a clear increasing or
decreasing trend (other than possibly a weak modulation in the amplitude of
complexity flucuations for the smallest scalar field amplitudes). Second,
we note that the complexity is 
near its maximum value at $t=0$ for initial scalar profiles of width
less than the AdS scale, while it is near its minimum at $t=0$ for widths
larger than the AdS scale. Finally, and strikingly, the fluctuation of the 
complexity as a function of origin time $t$ 
is nearly independent of the scalar field
amplitude $a$ until shortly before horizon formation. 
On the other hand, the time of the
dual field theory is the conformal boundary time $\tau$, which does not
scale in a simple way with $a$. Figure \ref{f:CVvstandtau} shows the complexity
for two sets of initial data plotted against origin time $t$ (subfigures
\ref{f:CVt-m0w21d5},\ref{f:CVt-m0w17d5}) and against conformal boundary time
$\tau$ (subfigures \ref{f:CVtau-m0w21d5},\ref{f:CVtau-m0w17d5}). As 
figure \ref{f:deltaspacetime} suggests, the $\C(\tau)$ curves are stretched 
compared to the $\C(t)$ curves, with stretching more pronounced shortly before
formation of a horizon. As a result, the curves for larger initial amplitudes
are stretched more at earlier times than the $\C(\tau)$ for smaller $a$,
and the complexity curves no longer align.

\begin{figure}[t]
\centering
\begin{subfigure}[t]{0.47\textwidth}
\includegraphics[width=\textwidth]{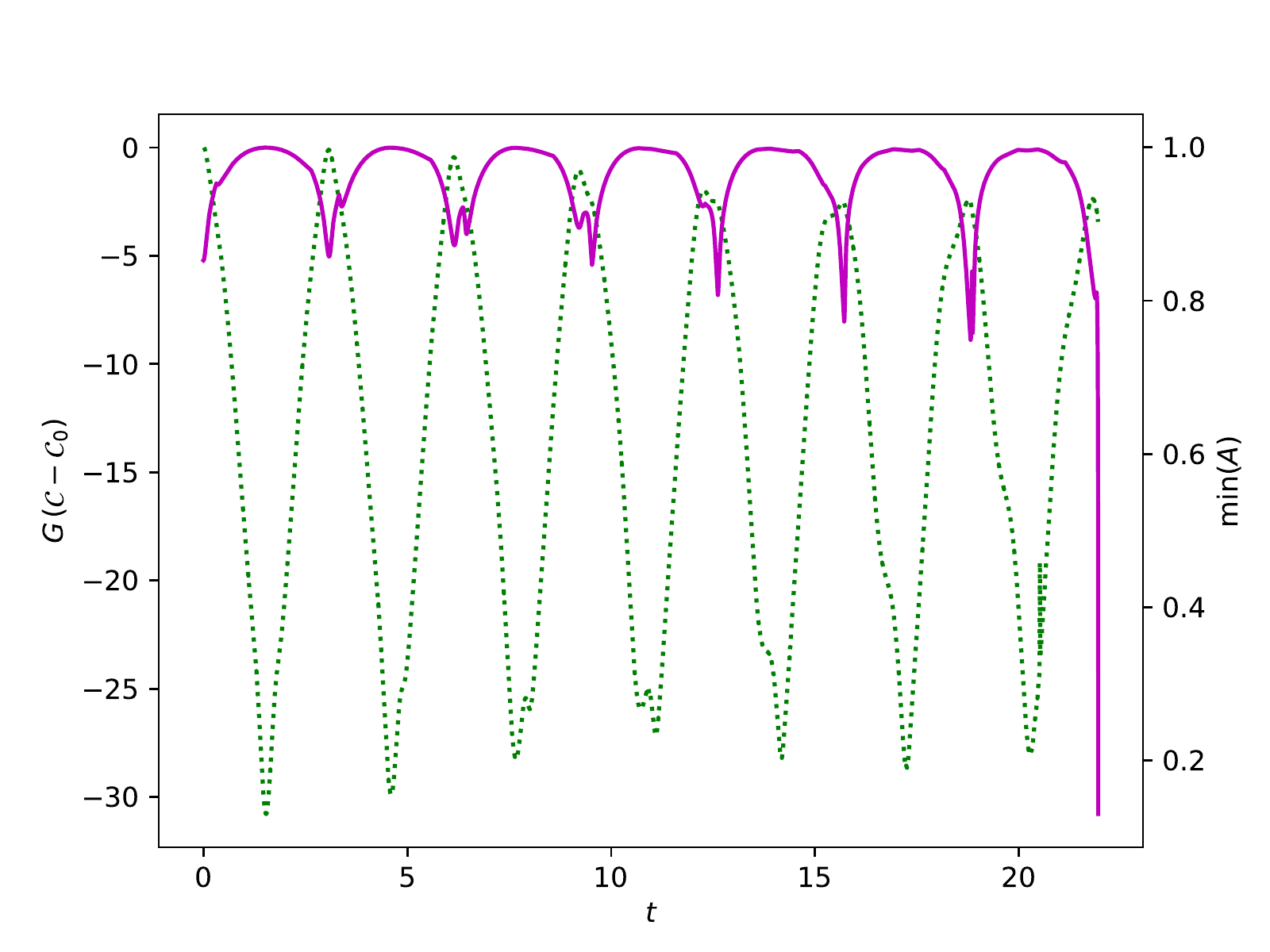}
\caption{$\sigma=0.25$, $a=5.0$}
\label{f:CVvsAmin-m0w025d4}\end{subfigure}
\begin{subfigure}[t]{0.47\textwidth}
\includegraphics[width=\textwidth]{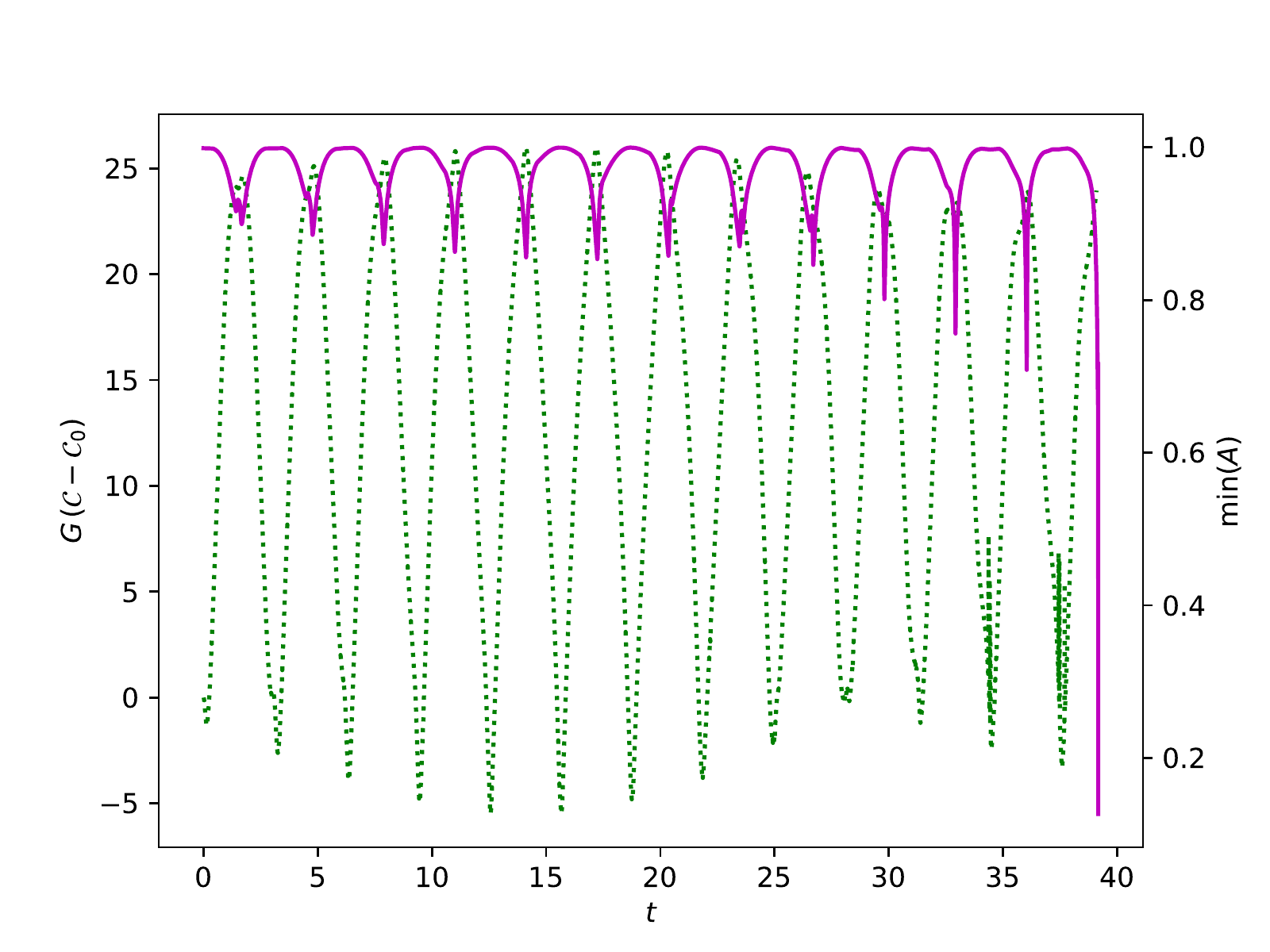}
\caption{$\sigma=8$, $a=0.10$}
\label{f:CVvsAmin-m0w8d4}\end{subfigure}
\caption{Volume complexity and minimum $A(x)$ vs $t$ for two sets of initial
data. Both are massless scalars in AdS$_4$. Complexity is dotted green and
minimum $A$ is solid magenta. The scales are shown on the left and right
vertical axes respectively.}
\label{f:CVvsAmin}
\end{figure}

We can understand the oscillatory behavior of the complexity in terms of the
complexity-momentum correspondence proposed in
\cite{Susskind:2018tei} and proved for volume complexity in 
spherically symmetric spacetimes 
in \cite{Barbon:2020olv}. In the evolutions presented here, the energy
of the scalar field is concentrated in thick shells (which may separate 
at times into thinner shells that nonetheless travel together). According
to the complexity-momentum correspondence, the complexity should increase 
as the shells travel inward (the momentum is infalling) and decrease 
as the shells travel outward. Therefore, the complexity should be at a local
maximum in time when the shells are closest to the origin.
To see that this happens here,
we redisplay the complexity of the smallest amplitude 
evolutions from figures \ref{f:CVt-m0w025d4} and \ref{f:CVt-m0w8d4}
in figure \ref{f:CVvsAmin}
alongside the minimum with respect to $x$ of the metric function $A(t,x)$.
We see that the local maxima of the complexity align with the local minima 
(in time) of $A$, which occur when the scalar pulse is most localized near
the origin, as expected.
The narrow initial data ($\sigma<1$) start at maximum complexity because
the energy of the scalar field is initially concentrated near the origin,
whereas wide initial data ($\sigma>1$) has energy initially in a shell
located at large radius, which then falls toward the origin. Note that
figure \ref{f:CVvsAmin} does not distinguish the contribution of $A$ from
that of $\delta$ to the complexity, since they are both most extreme at
the same times, though the fact that the maximal volume surface is nearly
a constant time surface hints that $A$ is more important.

It is also worth pondering the origin of the near amplitude independence of
the fluctuations of the complexity. From the gravitational
side of the duality, the rough scaling is relatively straightforward to 
understand
at least for $a\ll 1$: both $\delta$ and $A-1$ are $\O(a^2)$, so the time
dependence of the volume $V$ must also be $\O(a^2)$, so the change in
complexity is invariant for the $\ell=r_+$ prescription. 
However, this relationship is more mysterious
from the point of view of the dual field theory. Since the backgrounds are 
related by rescaling the scalar field and therefore the expectation value
of the dual operator $\O_\phi$, conformal invariance seems to play a role.
On first glance, that would seem also to explain why the time scale of
smaller amplitudes is somewhat stretched in boundary time $\tau$, leading to
alignment when plotted against origin time. This is problematic, though,
since $\tau-t\sim\O(a^2)$ at small amplitudes, but the near amplitude
invariance holds for massive as well as massless scalars, which correspond
to operators with different scaling dimensions. 
On the other hand, we could also consider the prescription with $\ell=1$, 
the AdS radius. Then the initial value $\C_0$ is (very close to) the
ground state complexity in each case. If the scalar field background 
corresponds to an $\O(a)$ perturbation of the CFT state that is 
perpendicular to the
vector from the reference state to the ground state (with regard to the
appropriate metric for complexity), $\C-\C_0$ will scale like $a^2$. Then
figure \ref{f:CVexamples} has the demonstrated amplitude invariance.
(This is a similar argument to that of 
\cite{arXiv:1806.08376,arXiv:1902.06499,arXiv:1903.04511}.)

If it holds as the amplitude vanishes, the scaling of complexity fluctuations
allows us to comment on the validity of Lloyd's bound, 
$|d\C/d\tau| \lesssim \M$, which \cite{Engelhardt:2021mju} proved recently
for volume complexity with $\ell=1$.\footnote{There is a weak additional 
mass dependence at large enough $\M$.} 
In particular, the validity of Lloyd's bound depends
on the reference scale $\ell$ chosen in the definition of $\C_V$.
First, in the $\ell=r_+$ prescription, $d\C/dt$ is unchanged as $a\to 0$,
and it becomes the left-hand-side of Lloyd's bound since $\tau\to t$ in 
that limit. On the other hand, $\M\to 0$, so Lloyd's bound must be
violated in the small amplitude limit. In contrast, for $\ell=1$, both
$d\C/d\tau$ and $\M$ scale as $a^2$, so our results are parametrically
consistent with Lloyd's bound. This result may be a point in favor of using
the AdS scale as the reference length in volume complexity.

The initial data shown in figures 
\ref{f:CVexamples},\ref{f:CVt-m0w21d5},\ref{f:CVtau-m0w21d5}
exhibit instability 
toward black hole formation at all computationally accessible amplitudes
(and also demonstrate the perturbative scaling $t_H\propto a^{-2}$ at small
enough amplitudes) \cite{arXiv:1711.00454}. 
We confirm that the time dependence of the
complexity has the same three behaviors for initial data that is stable
against horizon formation at low amplitudes in figures 
\ref{f:CVt-m0w17d5},\ref{f:CVtau-m0w17d5}. In these figures, the solid red curve
represents an amplitude that forms a horizon at $t_H\approx 8$, while the
dashed blue curve displays the complexity for a smaller amplitude that
does not form a horizon until at least $t=500$ (however, we display only 
until $t=35$ to reduce data storage requirements). In other words, prior to
horizon formation, the volume complexity is sensitive to eventual stability or
instability only through the difference in origin and conformal boundary
times.

As a final point, the reader may notice that $\C-\C_0$ can take negative 
values. On the other hand, \cite{Engelhardt:2021mju} proved recently that
volume complexity is always greater than the value for empty AdS, given a
few assumptions.\footnote{See \cite{Engelhardt:2021kyp} for one
exception to those assumptions.} This apparent discrepancy is due to the
contribution of the initial scalar field configuration to $\C_0$. We have
verified by numerical comparison that the minimum complexity for each 
of our evolutions is in fact greater than the complexity of empty AdS, so 
our results are consistent with \cite{Engelhardt:2021mju}.

\section{Complexity as Action}\label{s:action}
\subsection{Set-Up}\label{s:CAsetup}

The action complexity of the state on a fixed time slice $\Sigma$ of the
boundary is $\C_A=S_{WDW}/\pi$, where $S_{WDW}$ is the action of the 
Wheeler--DeWitt patch of $\Sigma$ in the bulk AdS spacetime, including terms
on the boundary of the WDW patch. Specifically, the WDW patch action is
$S_{WDW} = S + S_{f}+S_p +S_j$, where $S$ is the bulk action given in 
equation (\ref{action-bulk}), $S_{f,p}$ are boundary terms on the lightsheets
bounding the past and future of the WDW patch, and $S_j$ is a joint term
at the intersection of the two lightsheets. The lightsheet contributions are
\beq{action-lightsheet}
S_{f,p} = \pm \frac{1}{8\pi G} 
\int d\lambda d^{d-1}\theta \sqrt\gamma \left(\kappa
+\Theta\ln(\ell_{ct}\Theta)\vphantom{\frac 12}\right)
\eeq
over the future- or past-directed lightsheet emitted from $\Sigma$, where
$\lambda$ is a future-directed lightlike coordinate along the lightsheet, 
$\theta$ are 
angular coordinates on the lightsheet, $\gamma$ is the determinant of the
induced metric on the fixed-$\lambda$ slices,
$k^\mu$ is the normal (and therefore tangent or along $\lambda$) 
to the null hypersurface,
$\kappa$ can be defined by $k^\nu\Del_\nu k^\mu =\kappa k^\mu$, 
$\Theta\equiv \del_\lambda\ln \sqrt\gamma$ is the expansion of the lightsheet,
and $\ell_{ct}$ is an arbitrary lengthscale for the expansion counterterm.
The lightsheet and joint terms make the variational problem well-defined
within the WDW patch, as well as making the action parameterization invariant
\cite{arXiv:1609.00207}; \cite{arXiv:1804.07410} demonstrated the 
importance of the lightsheet expansion term to the action complexity.
The joint term is
\beq{action-joint}
S_j = -\frac{1}{8\pi G} \int d^{d-1}\theta\sqrt{\sigma} \left(\vphantom{\frac 12}
\ln(-k_f\cdot k_p/2) \right),
\eeq
where $\sigma$ is the determinant of the induced metric on the joint
and $k_{f,p}$ are the normals to the future- and past-directed
lightsheets intersecting at the joint. Since the scalar fields we consider
have $m^2\geq 0$, we do not include lightsheet or joint terms for the scalar
following \cite{arXiv:1903.04511,arXiv:2002.05779} 
(see also \cite{arXiv:1901.00014} for 
a discussion of boundary terms for scalars).

Like the volume complexity, the action diverges due to contributions near
the conformal boundary. To regulate it, we define the past and future
boundaries of the WDW patch as lightsheets joining at $t=t_0$,
$x=\pi/2-\epsilon$ for some small $\epsilon$. Then
we use the lightlike coordinates $\lambda=x$ on 
the past boundary of the WDW patch and $\lambda=\pi/2-x$ on the future
boundary, so $\lambda$ increases toward the future. Then the past and future
boundaries of the WDW patch are described by 
$dt_{p,f}/dx = \pm \exp(\delta)/A$ respectively. The on-shell bulk action is
therefore
\beq{bulk-onshell}
S = \frac{1}{16\pi G}\left( \frac{2\pi^{(d-1)/2}}{\Gamma(\frac{d-1}{2})}\right)
\int_0^{\pi/2-\epsilon} dx \int_{t_p(x)}^{t_f(x)} dt\, \tan^{d-1}(x)\sec^2(x)
e^{-\delta(t,x)}\left(-2d+2\mu^2 \phi^2\right) \eeq
with our choice of coordinates. The lightsheet contributions give
\begin{align}
S_f+S_p &=
\frac{1}{8\pi G}\left(\frac{2\pi^{(d-1)/2}}{\Gamma(\frac{d-1}{2})}\right)
\int_0^{\pi/2-\epsilon}dx\,\tan^{d-1}(x) \left[\del_x\delta(t_p(x),x)+
\del_x\delta(t_f(x),x)\vphantom{\frac{e^\delta}{A}}\right.\nonumber\\
&\left.+\left(\frac{e^\delta\del_t A}{A^2}\right)(t_p(x),x)
-\left(\frac{e^\delta\del_t A}{A^2}\right)(t_f(x),x)\right] .
\label{lightsheet-onshell}\end{align}
Note that the counterterms cancel for the two lightsheets because neither
$A$ nor $\delta$ enters into the angular metric (a consequence of spherical
symmetry). The joint term is
\beq{joint-onshell}
S_j = -\frac{1}{8\pi G}\left(\frac{2\pi^{(d-1)/2}}{\Gamma(\frac{d-1}{2})}\right)
\cot^{d-1}(\epsilon) \ln\left[\csc^2(\epsilon)/A\right] \approx
\frac{1}{8\pi G}\left(\frac{2\pi^{(d-1)/2}}{\Gamma(\frac{d-1}{2})}\right)
\epsilon^{1-d}\left[2\ln\epsilon+\ln A\right] .\eeq
The metric function $A$ has the expansion 
$A=1-2\M\epsilon^d/(d-1)+\O(\epsilon^{d+2})$ near $x=\pi/2$, so, while $S_j$
diverges for $\epsilon\to 0$, the time-dependence (and in fact the entire
deviation from empty AdS) vanishes in that limit.

\subsection{Methods and Results}\label{s:CAresults}

\begin{figure}[t]
\centering
\begin{subfigure}[t]{0.47\textwidth}
\includegraphics[width=\textwidth]{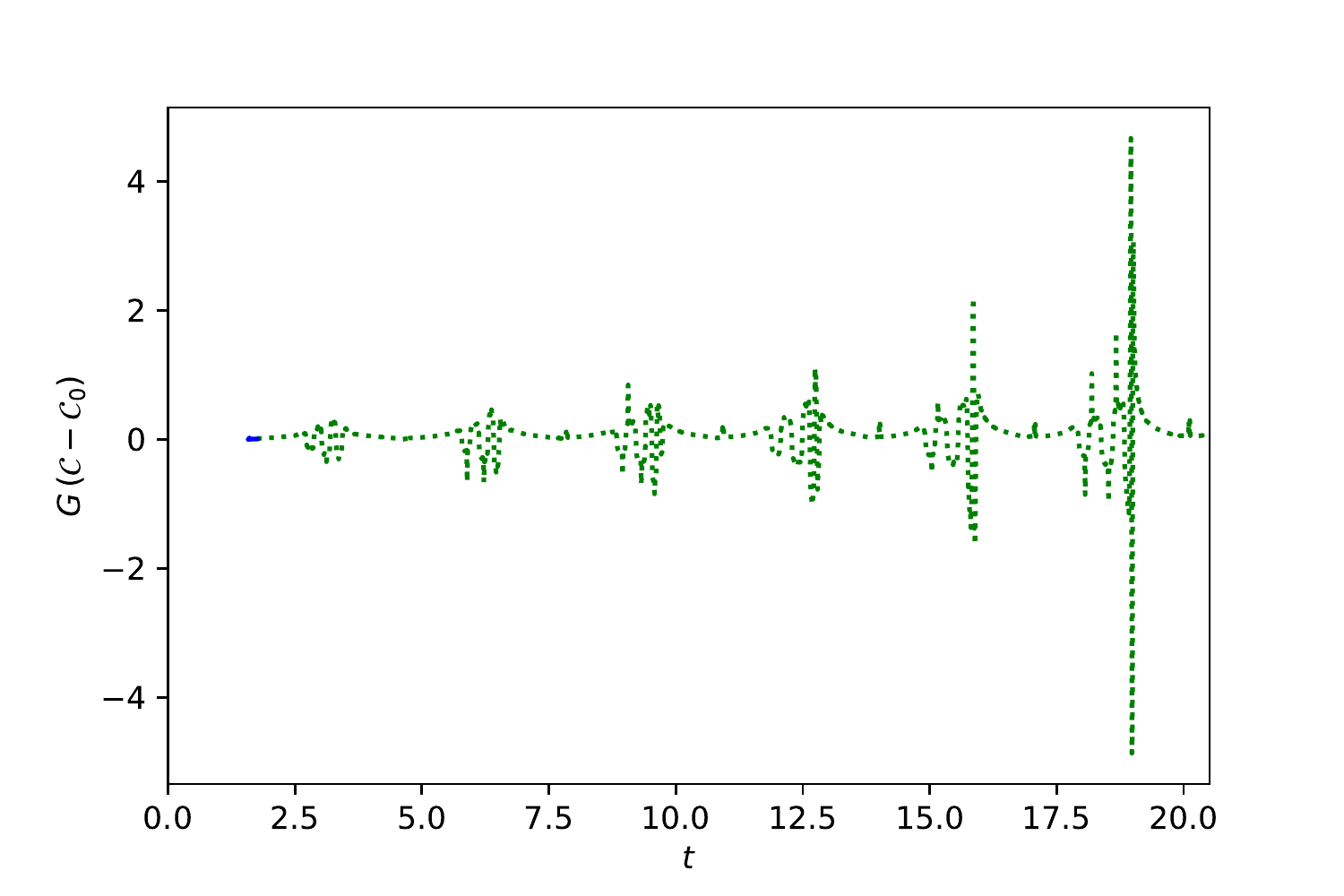}
\caption{AdS$_5$, $\mu=0$, $\sigma=0.25$, $a=6.39,3.87$}
\label{f:CAt-m0w025d5}\end{subfigure}
\begin{subfigure}[t]{0.47\textwidth}
\includegraphics[width=\textwidth]{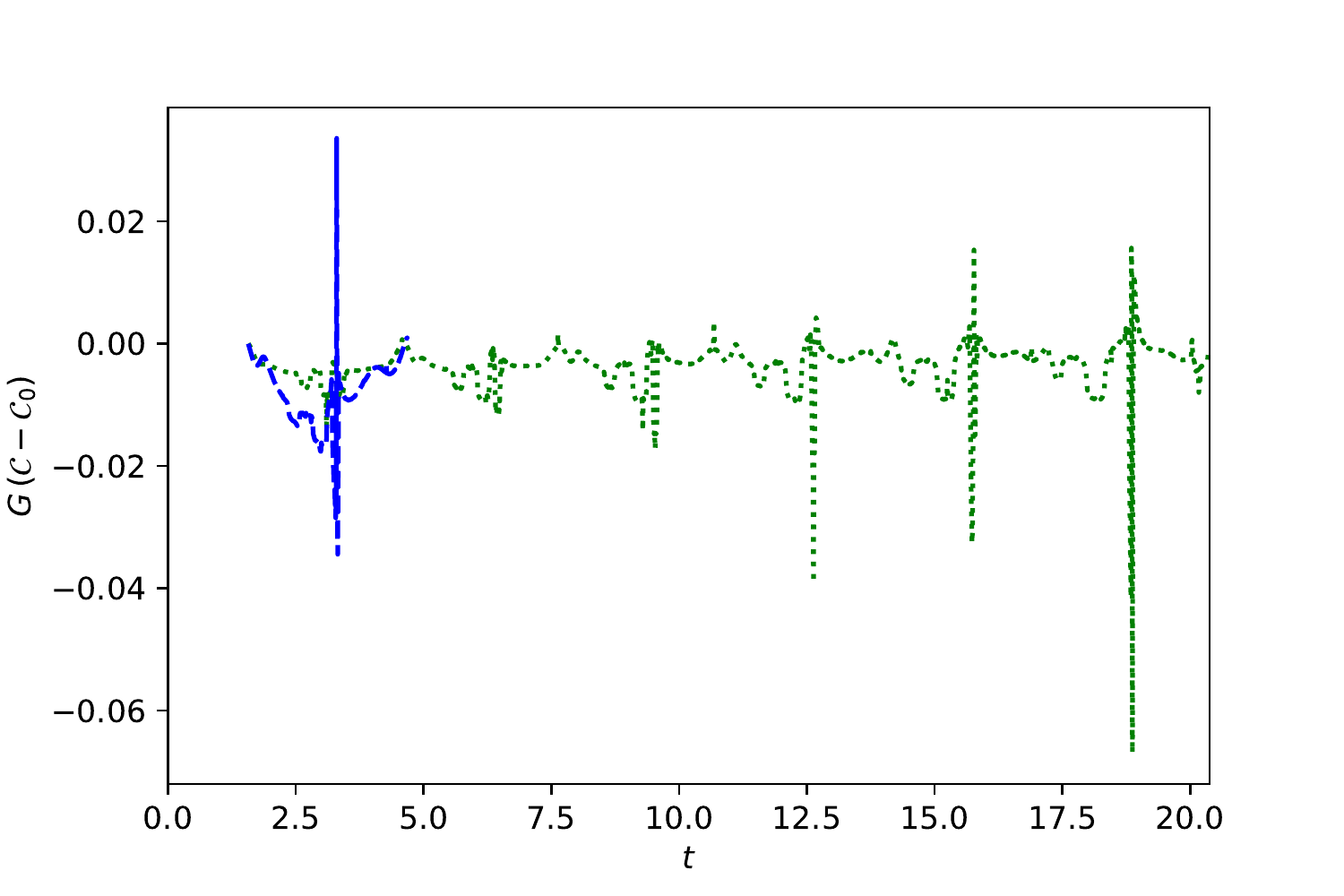}
\caption{AdS$_4$, $\mu=0$, $\sigma=0.25$, $a=15,8.0,5.0$}
\label{f:CAt-m0w025d4}\end{subfigure}
\begin{subfigure}[t]{0.47\textwidth}
\includegraphics[width=\textwidth]{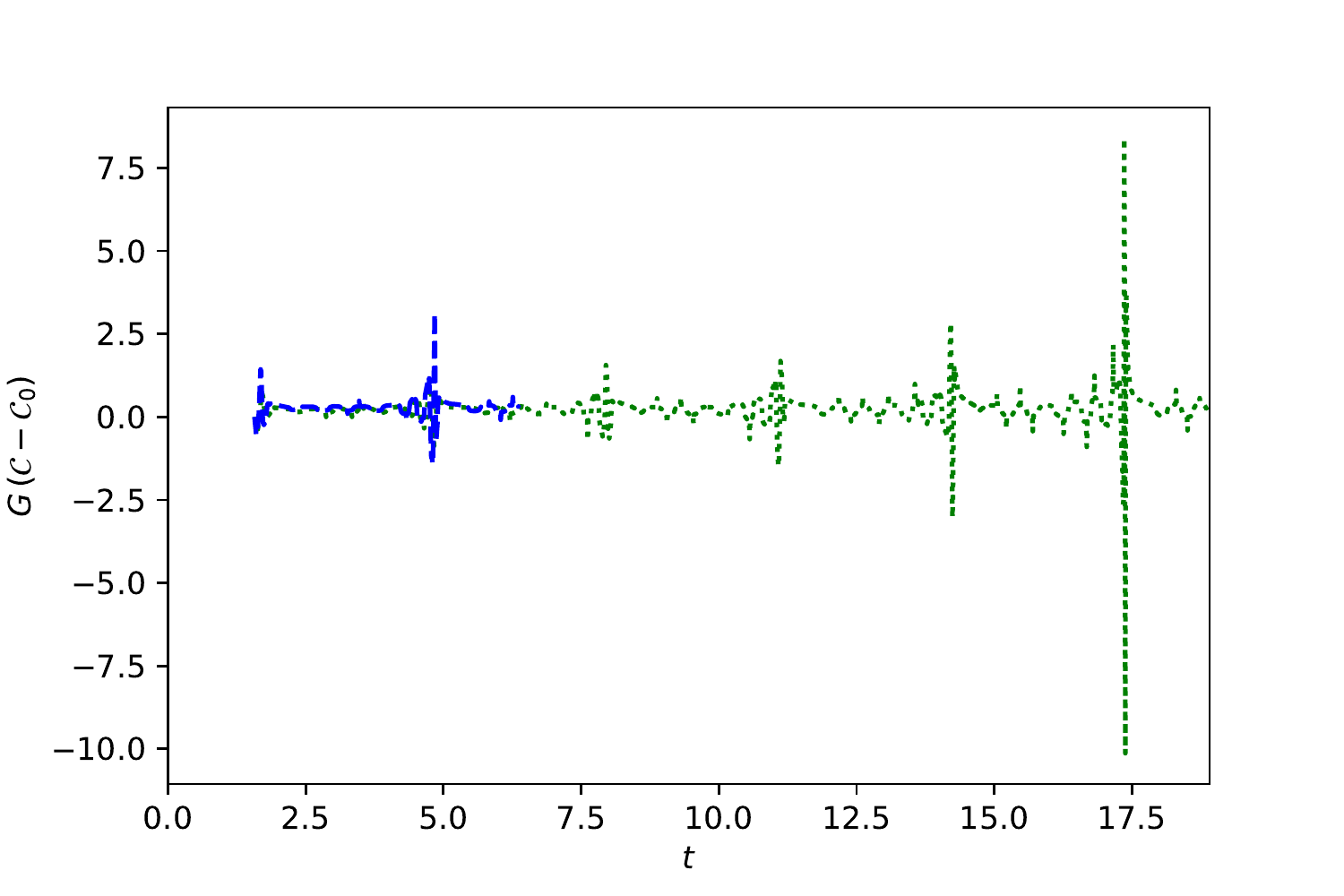}
\caption{AdS$_5$, $\mu=5$, $\sigma=2$, $a=0.12,0.10$}
\label{f:CAt-m5w2d5}\end{subfigure}
\begin{subfigure}[t]{0.47\textwidth}
\includegraphics[width=\textwidth]{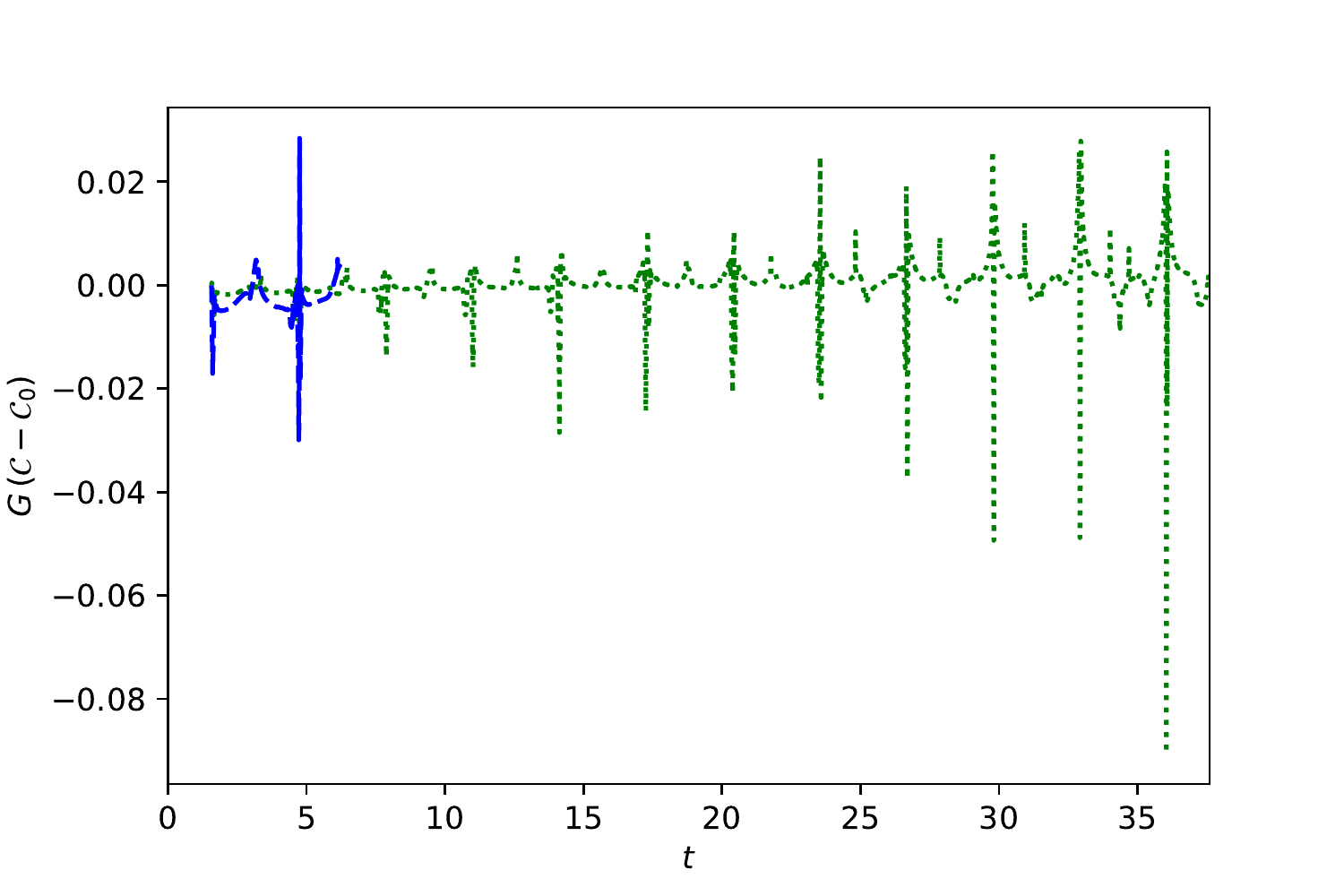}
\caption{AdS$_4$, $\mu=0$, $\sigma=8$, $a=0.14,0.10$}
\label{f:CAt-m0w8d4}\end{subfigure}
\begin{subfigure}[t]{0.47\textwidth}
\includegraphics[width=\textwidth]{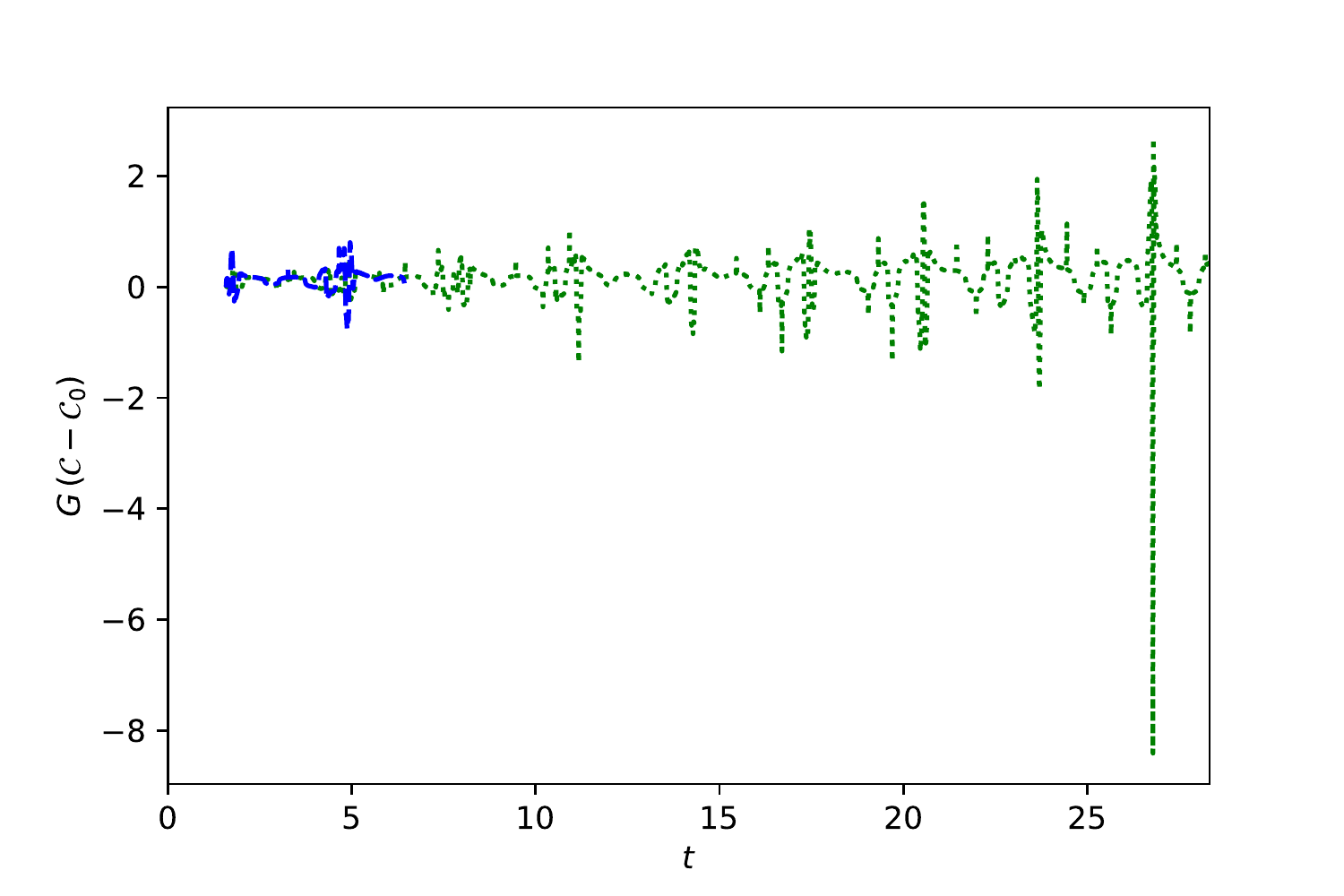}
\caption{AdS$_5$, $\mu=0$, $\sigma=2.1$, $a=0.24,0.18$}
\label{f:CAt-m0w21d5}\end{subfigure}
\begin{subfigure}[t]{0.47\textwidth}
\includegraphics[width=\textwidth]{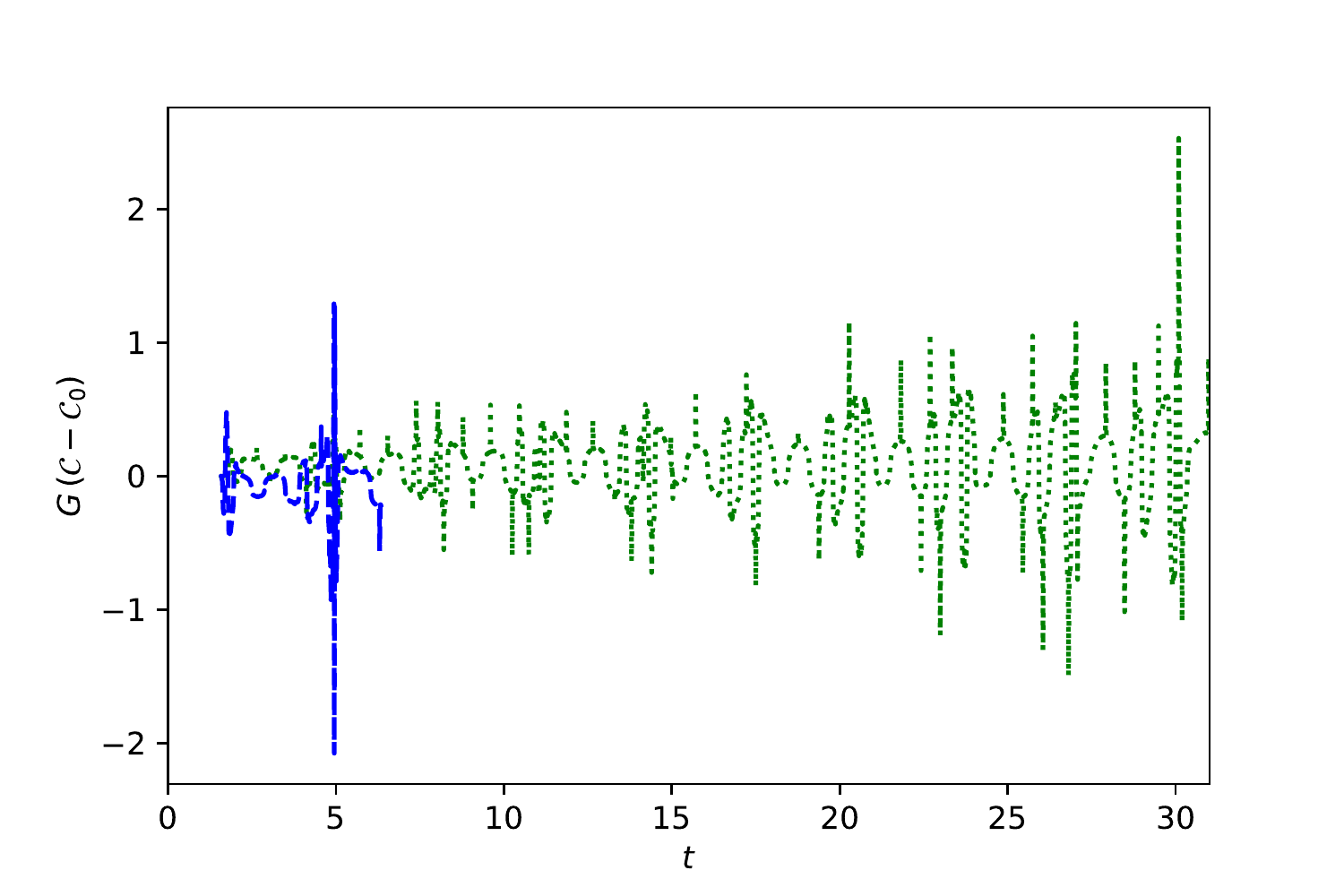}
\caption{AdS$_5$, $\mu=0$, $\sigma=1.7$, $a=0.38,0.22$}
\label{f:CAt-m0w17d5}\end{subfigure}
\caption{Rescaled action complexity for listed initial data in AdS$_4$ and 
AdS$_5$.
Curve for the larger amplitude in each figure is dashed blue, 
smaller is dotted green.} \label{f:CAexamples}
\end{figure}

To find the past and future boundaries $t_p(x),t_f(x)$ of the WDW patch,
we again use the scipy function odeint, and we carry out the integrals
(\ref{bulk-onshell},\ref{lightsheet-onshell}) 
using the quad and nquad integration
routines from scipy. To prevent errors in the bulk integral, we raised the
number of nquad subdivisions to $5\times 10^5$ from the default value of 50.
Just as for the volume complexity calculation, we interpolate the functions
$A,\delta$ as necessary and evaluate derivatives through differencing.
However, the complexity is evaluated at the time 
$t_0=t_p(\pi/2-\epsilon)=t_f(\pi/2-\epsilon)$ and 
$t_{p,f}(0)\approx t_0\pm\pi/2$, so we can only evaluate the complexity over
the approximate range $\pi/2\lesssim t_0\lesssim t_H-\pi/2$. For other times,
the WDW patch extends outside the region of spacetime we have simulated.

Figure \ref{f:CAexamples} shows the action complexity for the same sets of
initial data as figures \ref{f:CVexamples} and \ref{f:CVvstandtau}. Because
evaluating the complexity at time $t_0$ on the boundary cutoff surface
requires bulk data for times $t_0-\pi/2\leq t\leq t_0+\pi/2$, the curves
terminate at $t_0\approx t_H-\pi/2$.
As a result, we do not evaluate the action complexity for the largest
amplitude in each set of initial data when those form a horizon before 
$t=\pi/2$. Furthermore, since the behavior is similar (just stretched
horizontally) when plotted versus boundary time $\tau$, we show only the
$\C$ vs $t$ plots.

In each case, the complexity is roughly constant when averaged
over a period of the scalar field oscillations, though there are pronounced
spikes each period. Furthermore, while the complexity does not appear to 
have a trend on average, the spikes do increase in magnitude, especially 
near horizon formation. In fact, the spike size also increases
somewhat for the smaller amplitude in figure \ref{f:CAt-m0w17d5}, which is
a stable evolution. We have verified that the dominant contribution to the
time dependent complexity is the bulk integral and that the large spikes in
$\C$ are at times when the scalar field pulse is passing through the 
origin; the spikes are due to greater time dilation effects since $\delta$ is
most negative at those times. Therefore, we see that the progressive
focusing of the scalar pulse as seen in the time dilation function $\delta$ 
also leads to a ratcheting effect in the magnitude of the spikes in the
action complexity.

It is also worth noting that the action complexity for given width initial 
data has a similar shape for different initial scalar amplitudes; however,
the action complexity does not have the same amplitude independence that
we have seen for volume complexity. In fact, we have also checked whether 
this amplitude independence appears if we rescale the complexity by the
square amplitude, but it does not.

\section{Discussion}\label{s:discussion}

We have computed the holographic complexity of scalar field matter in the
process of gravitational collapse prior to (an approximate notion of) black
hole horizon formation using both $\C=V$ and $\C=A$ proposals. In contrast
to black hole formation by thin shells in AdS-Vaidya spacetimes, the matter
distributions we study are smooth and may not form a horizon until times
late compared to the AdS scale. However, because our numerical solutions
to the Klein-Gordon and Einstein equations are restricted to times before
horizon formation, we are limited to studying early time transients in the
complexity. In terms of the AdS/CFT correspondence, these backgrounds are
dual to thermalization of an initial energy distribution (strictly speaking,
we mean distribution of energy through modes of the CFT 
with a large range of scales). Therefore, our results indicate the time
dependence of complexity during the approach to equilibrium of a CFT.

In global AdS coordinates, a time slice of the boundary is a sphere, so
we might expect any energy in the CFT to equilibrate eventually because it
cannot dissipate. In fact, if the initial scalar field profile is either
wide or narrow compared to the AdS width, the scalar field matter appears 
to form a black hole even at arbitrarily small amplitudes. The intuition,
supported by the time evolution of some metric components, is that the
scalar field undergoes gravitational focusing as it oscillates in the AdS,
eventually collapsing. However, due to 
the high degree of symmetry of AdS, some initial scalar profiles are stable
against horizon formation at small amplitudes with quasiperiodic time
evolution. This dichotomy naturally leads us to ask if holographic 
complexity can provide a diagnostic of stability or instability against
collapse before the moment of horizon formation. The answer appears to be
in the negative, as neither the volume nor action complexity is qualitatively
different for stable versus unstable initial field profiles.
Another question of interest is whether complexity behaves quasiperiodically
or exhibits a ratcheting effect (some trend imposed on oscillatory behavior) 
due to focusing of the scalar field pulse.

The volume complexity appears quasiperiodic in time; the amplitude of 
oscillations in complexity is modulated (particularly for smaller scalar field
amplitudes), but there is not a clear trend. Further, the complexity starts
near its maximum or minimum for initial data that is narrower or wider than
the AdS scale respectively. Most strikingly, if we use the Schwarschild 
radius for the length scale $\ell$ in the definition of volume complexity,
the complexity (with initial value subtracted) 
is nearly invariant under change of amplitude; with $\ell$ equal the AdS 
radius, fluctuations in complexity scale as the second power of the scalar
field amplitude. The origin of this property on the gravitational side of
the duality is clear, but it is somewhat more obsure from the point of view
of the boundary CFT. It may be related to conformal invariance of the
dual field theory or to orthogonality between the ground state and the
fluctuation due to the scalar field. This scaling property means that
Lloyd's bound is violated if the reference scale is the Schwarzschild radius.

The action complexity is also quasiperiodic in nature, except for large 
spikes in $\C-\C_0$ once per period (with period $\Delta t\sim\pi$) and 
some smaller spikes in between. These large
spikes correspond to strong deviations of the time dilation function $\delta$,
which expand the portion of the WDW patch near the boundary. While the
overall complexity curve appears quasiperiodic, the large spikes increase
in magnitude, particularly as the system approaches horizon formation. 
That is, the action complexity demonstrates a ratcheting behavior, 
becoming more extreme as the scalar pulse undergoes gravitational focusing.

We close with a few words about the relation of this work to the discussion
of holographic complexity for coherent states in 
\cite{arXiv:1903.04511,arXiv:1912.10436,arXiv:2002.05779}. 
Coherent states of the CFT are
given by scalar fields in AdS, like we study here. We have considered
the nonperturbative evolution of those scalar fields, including backreaction
from gravity in AdS. In contrast, \cite{arXiv:1903.04511,arXiv:2002.05779}
studied perturbative scalar field amplitudes to second order, which 
captures the gravitational effects of the field but not the 
backreaction on the field itself. Therefore, they cannot study the 
gravitational collapse of the field leading to horizon formation.
On the other hand, the numerical demands
of recreating the spacetime geometry mean that we are unable to consider
perturbatively small scalar field amplitudes for long time intervals.
It would be interesting in the future to compare a numerical study of 
scalar field evolution to the perturbative results of
\cite{arXiv:1903.04511,arXiv:2002.05779}.

\acknowledgments
The authors wish to thank N.~Deppe for use of the gravitational collapse
code from \cite{arXiv:1410.1869,arXiv:1508.02709}, and ARF thanks Z.~Fisher
and \AA.~Folkestad
for helpful conversations. The work of ARF and MPG was supported by the
Natural Sciences and Engineering Research Council of Canada Discovery Grant
program, grants 2015-00046 and 2020-00054. MPG was additionally supported
by the Natural Sciences and Engineering Research Council of Canada USRA
program. The work of MS was supported by the Mitacs Globalink program.


\bibliographystyle{JHEP}
\bibliography{collapsecomplexity}

\providecommand{\href}[2]{#2}\begingroup\raggedright\begin{thebibliography}{10}

\bibitem{arXiv:1402.5674}
L.~Susskind, \emph{{Computational Complexity and Black Hole Horizons}},
  \href{https://doi.org/10.1002/prop.201500092}{\emph{Fortsch. Phys.}
  {\bfseries 64} (2016) 24} [\href{https://arxiv.org/abs/1403.5695}{{\ttfamily
  1403.5695}}].

\bibitem{arXiv:1406.2678}
D.~Stanford and L.~Susskind, \emph{{Complexity and Shock Wave Geometries}},
  \href{https://doi.org/10.1103/PhysRevD.90.126007}{\emph{Phys. Rev. D}
  {\bfseries 90} (2014) 126007}
  [\href{https://arxiv.org/abs/1406.2678}{{\ttfamily 1406.2678}}].

\bibitem{arXiv:1509.07876}
A.~R. Brown, D.~A. Roberts, L.~Susskind, B.~Swingle and Y.~Zhao,
  \emph{{Holographic Complexity Equals Bulk Action?}},
  \href{https://doi.org/10.1103/PhysRevLett.116.191301}{\emph{Phys. Rev. Lett.}
  {\bfseries 116} (2016) 191301}
  [\href{https://arxiv.org/abs/1509.07876}{{\ttfamily 1509.07876}}].

\bibitem{arXiv:1512.04993}
A.~R. Brown, D.~A. Roberts, L.~Susskind, B.~Swingle and Y.~Zhao,
  \emph{{Complexity, action, and black holes}},
  \href{https://doi.org/10.1103/PhysRevD.93.086006}{\emph{Phys. Rev. D}
  {\bfseries 93} (2016) 086006}
  [\href{https://arxiv.org/abs/1512.04993}{{\ttfamily 1512.04993}}].

\bibitem{quant-ph/0502070}
M.~A. Nielsen, \emph{A geometric approach to quantum circuit lower bounds},
  \href{https://arxiv.org/abs/quant-ph/0502070}{{\ttfamily quant-ph/0502070}}.

\bibitem{quant-ph/0603161}
M.~A. Nielsen, M.~R. Dowling, M.~Gu and A.~C. Doherty, \emph{Quantum
  computation as geometry},
  \href{https://doi.org/10.1126/science.1121541}{\emph{Science} {\bfseries 311}
  (2006) 1133–1135} [\href{https://arxiv.org/abs/quant-ph/0603161}{{\ttfamily
  quant-ph/0603161}}].

\bibitem{quant-ph/0701004}
M.~R. Dowling and M.~A. Nielsen, \emph{The geometry of quantum computation},
  \href{https://arxiv.org/abs/quant-ph/0701004}{{\ttfamily quant-ph/0701004}}.

\bibitem{arXiv:1707.08570}
R.~Jefferson and R.~C. Myers, \emph{{Circuit complexity in quantum field
  theory}}, \href{https://doi.org/10.1007/JHEP10(2017)107}{\emph{JHEP}
  {\bfseries 10} (2017) 107}
  [\href{https://arxiv.org/abs/1707.08570}{{\ttfamily 1707.08570}}].

\bibitem{arXiv:1801.07620}
R.~Khan, C.~Krishnan and S.~Sharma, \emph{{Circuit Complexity in Fermionic
  Field Theory}}, \href{https://doi.org/10.1103/PhysRevD.98.126001}{\emph{Phys.
  Rev. D} {\bfseries 98} (2018) 126001}
  [\href{https://arxiv.org/abs/1801.07620}{{\ttfamily 1801.07620}}].

\bibitem{arXiv:1803.10638}
L.~Hackl and R.~C. Myers, \emph{{Circuit complexity for free fermions}},
  \href{https://doi.org/10.1007/JHEP07(2018)139}{\emph{JHEP} {\bfseries 07}
  (2018) 139} [\href{https://arxiv.org/abs/1803.10638}{{\ttfamily
  1803.10638}}].

\bibitem{arXiv:1408.2823}
L.~Susskind and Y.~Zhao, \emph{{Switchbacks and the Bridge to Nowhere}},
  \href{https://arxiv.org/abs/1408.2823}{{\ttfamily 1408.2823}}.

\bibitem{arXiv:1711.02668}
M.~Moosa, \emph{{Evolution of Complexity Following a Global Quench}},
  \href{https://doi.org/10.1007/JHEP03(2018)031}{\emph{JHEP} {\bfseries 03}
  (2018) 031} [\href{https://arxiv.org/abs/1711.02668}{{\ttfamily
  1711.02668}}].

\bibitem{arXiv:1802.06740}
M.~Alishahiha, A.~Faraji~Astaneh, M.~R. Mohammadi~Mozaffar and A.~Mollabashi,
  \emph{{Complexity Growth with Lifshitz Scaling and Hyperscaling Violation}},
  \href{https://doi.org/10.1007/JHEP07(2018)042}{\emph{JHEP} {\bfseries 07}
  (2018) 042} [\href{https://arxiv.org/abs/1802.06740}{{\ttfamily
  1802.06740}}].

\bibitem{arXiv:1803.11162}
D.~S. Ageev, I.~Y. Aref'eva, A.~A. Bagrov and M.~I. Katsnelson,
  \emph{{Holographic local quench and effective complexity}},
  \href{https://doi.org/10.1007/JHEP08(2018)071}{\emph{JHEP} {\bfseries 08}
  (2018) 071} [\href{https://arxiv.org/abs/1803.11162}{{\ttfamily
  1803.11162}}].

\bibitem{arXiv:1804.07410}
S.~Chapman, H.~Marrochio and R.~C. Myers, \emph{{Holographic complexity in
  Vaidya spacetimes. Part I}},
  \href{https://doi.org/10.1007/JHEP06(2018)046}{\emph{JHEP} {\bfseries 06}
  (2018) 046} [\href{https://arxiv.org/abs/1804.07410}{{\ttfamily
  1804.07410}}].

\bibitem{Lezgi:2021qog}
M.~Lezgi and M.~Ali-Akbari, \emph{{Complexity and uncomplexity during energy
  injection}}, \href{https://doi.org/10.1103/PhysRevD.103.126024}{\emph{Phys.
  Rev. D} {\bfseries 103} (2021) 126024}
  [\href{https://arxiv.org/abs/2103.05023}{{\ttfamily 2103.05023}}].

\bibitem{arXiv:1104.3702}
P.~Bizon and A.~Rostworowski, \emph{{On weakly turbulent instability of anti-de
  Sitter space}},
  \href{https://doi.org/10.1103/PhysRevLett.107.031102}{\emph{Phys. Rev. Lett.}
  {\bfseries 107} (2011) 031102}
  [\href{https://arxiv.org/abs/1104.3702}{{\ttfamily 1104.3702}}].

\bibitem{arXiv:1106.2339}
D.~Garfinkle and L.~A. Pando~Zayas, \emph{{Rapid Thermalization in Field Theory
  from Gravitational Collapse}},
  \href{https://doi.org/10.1103/PhysRevD.84.066006}{\emph{Phys. Rev. D}
  {\bfseries 84} (2011) 066006}
  [\href{https://arxiv.org/abs/1106.2339}{{\ttfamily 1106.2339}}].

\bibitem{arXiv:1108.4539}
J.~Jalmuzna, A.~Rostworowski and P.~Bizon, \emph{{A Comment on AdS collapse of
  a scalar field in higher dimensions}},
  \href{https://doi.org/10.1103/PhysRevD.84.085021}{\emph{Phys. Rev. D}
  {\bfseries 84} (2011) 085021}
  [\href{https://arxiv.org/abs/1108.4539}{{\ttfamily 1108.4539}}].

\bibitem{arXiv:1110.5823}
D.~Garfinkle, L.~A. Pando~Zayas and D.~Reichmann, \emph{{On Field Theory
  Thermalization from Gravitational Collapse}},
  \href{https://doi.org/10.1007/JHEP02(2012)119}{\emph{JHEP} {\bfseries 02}
  (2012) 119} [\href{https://arxiv.org/abs/1110.5823}{{\ttfamily 1110.5823}}].

\bibitem{arXiv:1304.4166}
A.~Buchel, S.~L. Liebling and L.~Lehner, \emph{{Boson stars in AdS spacetime}},
  \href{https://doi.org/10.1103/PhysRevD.87.123006}{\emph{Phys. Rev. D}
  {\bfseries 87} (2013) 123006}
  [\href{https://arxiv.org/abs/1304.4166}{{\ttfamily 1304.4166}}].

\bibitem{arXiv:1307.2875}
M.~Maliborski and A.~Rostworowski, \emph{{A comment on ''Boson stars in
  AdS''}},  \href{https://arxiv.org/abs/1307.2875}{{\ttfamily 1307.2875}}.

\bibitem{arXiv:1308.1235}
M.~Maliborski and A.~Rostworowski, \emph{{Lecture Notes on Turbulent
  Instability of Anti-de Sitter Spacetime}},
  \href{https://doi.org/10.1142/S0217751X13400204}{\emph{Int. J. Mod. Phys. A}
  {\bfseries 28} (2013) 1340020}
  [\href{https://arxiv.org/abs/1308.1235}{{\ttfamily 1308.1235}}].

\bibitem{Evnin:2021buq}
O.~Evnin, \emph{{Resonant Hamiltonian systems and weakly nonlinear dynamics in
  AdS spacetimes}},
  \href{https://doi.org/10.1088/1361-6382/ac1b46}{\emph{Class. Quant. Grav.}
  {\bfseries 38} (2021) 203001}
  [\href{https://arxiv.org/abs/2104.09797}{{\ttfamily 2104.09797}}].

\bibitem{arXiv:1504.05203}
H.~Okawa, J.~C. Lopes and V.~Cardoso, \emph{{Collapse of massive fields in
  anti-de Sitter spacetime}},
  \href{https://arxiv.org/abs/1504.05203}{{\ttfamily 1504.05203}}.

\bibitem{arXiv:1508.02709}
N.~Deppe and A.~R. Frey, \emph{{Classes of Stable Initial Data for Massless and
  Massive Scalars in Anti-de Sitter Spacetime}},
  \href{https://doi.org/10.1007/JHEP12(2015)004}{\emph{JHEP} {\bfseries 12}
  (2015) 004} [\href{https://arxiv.org/abs/1508.02709}{{\ttfamily
  1508.02709}}].

\bibitem{arXiv:1711.00454}
B.~Cownden, N.~Deppe and A.~R. Frey, \emph{{Phase diagram of stability for
  massive scalars in anti\textendash{}de Sitter spacetime}},
  \href{https://doi.org/10.1103/PhysRevD.102.026015}{\emph{Phys. Rev. D}
  {\bfseries 102} (2020) 026015}
  [\href{https://arxiv.org/abs/1711.00454}{{\ttfamily 1711.00454}}].

\bibitem{arXiv:1608.05402}
N.~Deppe, A.~Kolly, A.~R. Frey and G.~Kunstatter, \emph{{Black Hole Formation
  in AdS Einstein-Gauss-Bonnet Gravity}},
  \href{https://doi.org/10.1007/JHEP10(2016)087}{\emph{JHEP} {\bfseries 10}
  (2016) 087} [\href{https://arxiv.org/abs/1608.05402}{{\ttfamily
  1608.05402}}].

\bibitem{arXiv:1403.6471}
V.~Balasubramanian, A.~Buchel, S.~R. Green, L.~Lehner and S.~L. Liebling,
  \emph{{Holographic Thermalization, Stability of Anti\textendash{}de Sitter
  Space, and the Fermi-Pasta-Ulam Paradox}},
  \href{https://doi.org/10.1103/PhysRevLett.113.071601}{\emph{Phys. Rev. Lett.}
  {\bfseries 113} (2014) 071601}
  [\href{https://arxiv.org/abs/1403.6471}{{\ttfamily 1403.6471}}].

\bibitem{arXiv:1410.2631}
P.~Bizo\'n and A.~Rostworowski, \emph{{Comment on
  \textquotedblleft{}Holographic Thermalization, Stability of
  Anti\textendash{}de Sitter Space, and the Fermi-Pasta-Ulam
  Paradox\textquotedblright{}}},
  \href{https://doi.org/10.1103/PhysRevLett.115.049101}{\emph{Phys. Rev. Lett.}
  {\bfseries 115} (2015) 049101}
  [\href{https://arxiv.org/abs/1410.2631}{{\ttfamily 1410.2631}}].

\bibitem{arXiv:1506.07907}
V.~Balasubramanian, A.~Buchel, S.~R. Green, L.~Lehner and S.~L. Liebling,
  \emph{{Reply to Comment on \textquotedblleft{}Holographic Thermalization,
  Stability of Anti\textendash{}de Sitter Space, and the Fermi-Pasta-Ulam
  Paradox\textquotedblright{}}},
  \href{https://doi.org/10.1103/PhysRevLett.115.049102}{\emph{Phys. Rev. Lett.}
  {\bfseries 115} (2015) 049102}
  [\href{https://arxiv.org/abs/1506.07907}{{\ttfamily 1506.07907}}].

\bibitem{arXiv:1412.3249}
B.~Craps, O.~Evnin and J.~Vanhoof, \emph{{Renormalization, averaging,
  conservation laws and AdS (in)stability}},
  \href{https://doi.org/10.1007/JHEP01(2015)108}{\emph{JHEP} {\bfseries 01}
  (2015) 108} [\href{https://arxiv.org/abs/1412.3249}{{\ttfamily 1412.3249}}].

\bibitem{arXiv:1412.4761}
A.~Buchel, S.~R. Green, L.~Lehner and S.~L. Liebling, \emph{{Conserved
  quantities and dual turbulent cascades in anti\textendash{}de Sitter
  spacetime}}, \href{https://doi.org/10.1103/PhysRevD.91.064026}{\emph{Phys.
  Rev. D} {\bfseries 91} (2015) 064026}
  [\href{https://arxiv.org/abs/1412.4761}{{\ttfamily 1412.4761}}].

\bibitem{arXiv:1407.6273}
B.~Craps, O.~Evnin and J.~Vanhoof, \emph{{Renormalization group, secular term
  resummation and AdS (in)stability}},
  \href{https://doi.org/10.1007/JHEP10(2014)048}{\emph{JHEP} {\bfseries 10}
  (2014) 048} [\href{https://arxiv.org/abs/1407.6273}{{\ttfamily 1407.6273}}].

\bibitem{arXiv:1612.07701}
P.~Carracedo, J.~Mas, D.~Musso and A.~Serantes, \emph{{Adiabatic pumping
  solutions in global AdS}},
  \href{https://doi.org/10.1007/JHEP05(2017)141}{\emph{JHEP} {\bfseries 05}
  (2017) 141} [\href{https://arxiv.org/abs/1612.07701}{{\ttfamily
  1612.07701}}].

\bibitem{arXiv:1712.07637}
A.~Biasi, P.~Carracedo, J.~Mas, D.~Musso and A.~Serantes, \emph{{Floquet Scalar
  Dynamics in Global AdS}},
  \href{https://doi.org/10.1007/JHEP04(2018)137}{\emph{JHEP} {\bfseries 04}
  (2018) 137} [\href{https://arxiv.org/abs/1712.07637}{{\ttfamily
  1712.07637}}].

\bibitem{arXiv:1912.07143}
B.~Cownden, \emph{{Examining Instabilities Due to Driven Scalars in AdS}},
  \href{https://doi.org/10.1007/JHEP12(2020)013}{\emph{JHEP} {\bfseries 12}
  (2020) 013} [\href{https://arxiv.org/abs/1912.07143}{{\ttfamily
  1912.07143}}].

\bibitem{arXiv:1410.1869}
N.~Deppe, A.~Kolly, A.~Frey and G.~Kunstatter, \emph{{Stability of AdS in
  Einstein Gauss Bonnet Gravity}},
  \href{https://doi.org/10.1103/PhysRevLett.114.071102}{\emph{Phys. Rev. Lett.}
  {\bfseries 114} (2015) 071102}
  [\href{https://arxiv.org/abs/1410.1869}{{\ttfamily 1410.1869}}].

\bibitem{deppefreyhoult}
N.~Deppe, A.~R. Frey and R.~E. Hoult, \emph{work in progress}, .

\bibitem{arXiv:1807.02186}
J.~Couch, S.~Eccles, T.~Jacobson and P.~Nguyen, \emph{{Holographic Complexity
  and Volume}}, \href{https://doi.org/10.1007/JHEP11(2018)044}{\emph{JHEP}
  {\bfseries 11} (2018) 044}
  [\href{https://arxiv.org/abs/1807.02186}{{\ttfamily 1807.02186}}].

\bibitem{Susskind:2018tei}
L.~Susskind, \emph{{Why do Things Fall?}},
  \href{https://arxiv.org/abs/1802.01198}{{\ttfamily 1802.01198}}.

\bibitem{Barbon:2020olv}
J.~L.~F. Barbon, J.~Martin-Garcia and M.~Sasieta, \emph{{Proof of a
  Momentum/Complexity Correspondence}},
  \href{https://doi.org/10.1103/PhysRevD.102.101901}{\emph{Phys. Rev. D}
  {\bfseries 102} (2020) 101901}
  [\href{https://arxiv.org/abs/2006.06607}{{\ttfamily 2006.06607}}].

\bibitem{arXiv:1806.08376}
M.~Flory and N.~Miekley, \emph{{Complexity change under conformal
  transformations in AdS$_{3}$/CFT$_{2}$}},
  \href{https://doi.org/10.1007/JHEP05(2019)003}{\emph{JHEP} {\bfseries 05}
  (2019) 003} [\href{https://arxiv.org/abs/1806.08376}{{\ttfamily
  1806.08376}}].

\bibitem{arXiv:1902.06499}
M.~Flory, \emph{{WdW-patches in AdS$_{3}$ and complexity change under conformal
  transformations II}},
  \href{https://doi.org/10.1007/JHEP05(2019)086}{\emph{JHEP} {\bfseries 05}
  (2019) 086} [\href{https://arxiv.org/abs/1902.06499}{{\ttfamily
  1902.06499}}].

\bibitem{arXiv:1903.04511}
A.~Bernamonti, F.~Galli, J.~Hernandez, R.~C. Myers, S.-M. Ruan and J.~Sim\'on,
  \emph{{First Law of Holographic Complexity}},
  \href{https://doi.org/10.1103/PhysRevLett.123.081601}{\emph{Phys. Rev. Lett.}
  {\bfseries 123} (2019) 081601}
  [\href{https://arxiv.org/abs/1903.04511}{{\ttfamily 1903.04511}}].

\bibitem{Engelhardt:2021mju}
N.~Engelhardt and {\AA}.~Folkestad, \emph{{General Bounds on Holographic
  Complexity}},  \href{https://arxiv.org/abs/2109.06883}{{\ttfamily
  2109.06883}}.

\bibitem{Engelhardt:2021kyp}
N.~Engelhardt and {\AA}.~Folkestad, \emph{{Negative Complexity of Formation:
  the Compact Dimensions Strike Back}},
  \href{https://arxiv.org/abs/2111.14897}{{\ttfamily 2111.14897}}.

\bibitem{arXiv:1609.00207}
L.~Lehner, R.~C. Myers, E.~Poisson and R.~D. Sorkin, \emph{{Gravitational
  action with null boundaries}},
  \href{https://doi.org/10.1103/PhysRevD.94.084046}{\emph{Phys. Rev. D}
  {\bfseries 94} (2016) 084046}
  [\href{https://arxiv.org/abs/1609.00207}{{\ttfamily 1609.00207}}].

\bibitem{arXiv:2002.05779}
A.~Bernamonti, F.~Galli, J.~Hernandez, R.~C. Myers, S.-M. Ruan and J.~Sim\'on,
  \emph{{Aspects of The First Law of Complexity}},
  \href{https://arxiv.org/abs/2002.05779}{{\ttfamily 2002.05779}}.

\bibitem{arXiv:1901.00014}
K.~Goto, H.~Marrochio, R.~C. Myers, L.~Queimada and B.~Yoshida,
  \emph{{Holographic Complexity Equals Which Action?}},
  \href{https://doi.org/10.1007/JHEP02(2019)160}{\emph{JHEP} {\bfseries 02}
  (2019) 160} [\href{https://arxiv.org/abs/1901.00014}{{\ttfamily
  1901.00014}}].

\bibitem{arXiv:1912.10436}
S.~S. Hashemi, G.~Jafari and A.~Naseh, \emph{{First law of holographic
  complexity}}, \href{https://doi.org/10.1103/PhysRevD.102.106008}{\emph{Phys.
  Rev. D} {\bfseries 102} (2020) 106008}
  [\href{https://arxiv.org/abs/1912.10436}{{\ttfamily 1912.10436}}].

\end{thebibliography}\endgroup
\end{document}